\documentclass{aa} 
\usepackage{graphicx}
\usepackage{txfonts}
\usepackage{xcolor}
\definecolor{dblue}{rgb}{0.0, 0.0, 0.55}
\usepackage[colorlinks=true,urlcolor=dblue,citecolor=dblue,linkcolor=dblue]{hyperref}
\usepackage{mhchem}
\usepackage[normalem]{ulem}

\def\aap{A\&A}

\def\apjl{ApJL}
\def\apjs{ApJS}

\def\ao{Applied Optics}

\def\la{\mathrel{\mathchoice {\vcenter{\offinterlineskip\halign{\hfil
$\displaystyle##$\hfil\cr<\cr\sim\cr}}}
{\vcenter{\offinterlineskip\halign{\hfil$\textstyle##$\hfil\cr
<\cr\sim\cr}}}
{\vcenter{\offinterlineskip\halign{\hfil$\scriptstyle##$\hfil\cr
<\cr\sim\cr}}}
{\vcenter{\offinterlineskip\halign{\hfil$\scriptscriptstyle##$\hfil\cr
<\cr\sim\cr}}}}}

\def\Tg{T_{\rm g}}
\def\Td{T_{\rm d}}
\def\nH{n_{\langle\rm H\rangle}}

\def\St{S\hspace*{-0.3ex}t\hspace*{0.1ex}}

\def\Mdot{\dot{M}_{\rm acc}}
\def\kRoss{\kappa_{\rm R}}

\def\div{{\rm div}}
\def\grad{{\bf grad}}
\def\half{\hspace*{0.2pt}1/2\hspace*{-0.2pt}}

\def\amin{{a_{\rm min}}}
\def\amax{{a_{\rm max}}}
\def\apow{{a_{\rm pow}}}
\def\aset{{a_{\rm settle}}}
\def\neff{{n_{\rm eff}}}
\def\keff{{k_{\rm eff}}}
\def\permil{\ensuremath{{}^\text{o}\mkern-5mu/\mkern-3mu_\text{oo}}}

\definecolor{darkgreen}{rgb}{0.0, 0.5, 0.0}

\begin{document} 

\title{CAI formation in the early Solar System}

\author{P.~Woitke\inst{1}
        \and
        J.~Dr\k{a}\.zkowska\inst{2}
        \and
        H.~Lammer\inst{1}
        \and
        K.~Kadam\inst{1}
        \and
        P.~Marigo\inst{3}
}

\institute{Space Research Institute, Austrian Academy of Sciences,
           Schmiedlstr.~6, A-8042, Graz, Austria
           \and
           Max Planck Institute for Solar System Research,
           Justus-von-Liebig-Weg 3, 37077 G\"ottingen, Germany
           \and
           Department of Physics and Astronomy G.~Galilei,
           University of Padova, Vicolo dell'Osservatorio 3,
           I-35122, Padova, Italy
}

\date{Received 8 April 2024 / Accepted 23 April 2024}

\abstract{Ca-Al-rich inclusions (CAIs) are the oldest dated solid materials in the solar system, found as light-coloured crystalline ingredients in carbonaceous chondrite meteorites. Their formation time is commonly associated with age zero of the Solar System. Yet, the physical and chemical processes that once led to the formation of these sub-millimetre to centimetre-sized mineral particles in the early solar nebula are still a matter of debate. This paper proposes a pathway to form such inclusions during the earliest phases of disc evolution. We combine 1D viscous disc evolutionary models with 2D radiative transfer, equilibrium condensation, and new dust opacity calculations. We show that the viscous heating associated with the high accretion rates in the earliest evolutionary phases causes the midplane inside of about 0.5\,au to heat up to limiting temperatures of about 1500-1700\,K, but no further. These high temperatures force all refractory material components of the inherited interstellar dust grains to sublimate -- except for a few Al-Ca-Ti oxides such as Al$_2$O$_3$, Ca$_2$Al$_2$SiO$_7$, and CaTiO$_3$. This is a recurring and very stable result in all our simulations since these minerals form a natural thermostat. Once the Mg-Fe silicates are gone, the dust becomes more transparent and the heat is more efficiently transported to the disc surface, which prevents any further warm-up. This thermostat mechanism keeps these minerals above their annealing temperature for hundreds of thousands of years, allowing them to form large, pure and crystalline particles. These particles are dragged out by the viscously spreading disc, and once they reach a distance of about 0.5\,au, the silicates re-condense on the surface of the Ca-Al-rich particles, adding an amorphous silicate matrix.  We estimate that this mechanism to produce CAIs works during the first 50\,000 years of disc evolution. These particles then continue to move outward and populate the entire disc up to radii of about 50\,au, before, eventually, the accretion rate subsides, the disc cools, and the particles start to drift inwards.}

\maketitle


\section{Introduction}
New stars are born when the cold cores of molecular clouds collapse under the pull of their own gravity.
Due to the conservation of angular momentum during cloud collapse, a rapidly rotating disc is formed around the central protostar \citep{Shu1987,Williams2011}.
The chemical composition of this proto-planetary disc, and the physical processes occurring therein, determine the elemental and molecular environment in which new planetary systems form \citep{Oberg2023}. 

Important information about the environmental conditions in the early Solar System, before the planets formed, can be deduced from the study of the material components found in primitive meteorites (chondrites): the matrix, the silicate-rich spherical chondrules, and the Ca-Al-rich inclusions \citep[CAIs,][]{Kita2012,Kruijer2017}. 

The matrix is the background material into which the other constituents like the chondrules, the CAIs, and larger mineral grains are embedded \citep{Brearley1989}. The matrix material consists mainly of anhydrous meteoritic phases, such as olivine, pyroxene, and Fe-Ni metals that have been produced by transient heating and rapid cooling about $2-4$\,Myrs after the CAIs originated. Lesser amounts of sulphides, sulphates, carbonates, and several other minerals are also found, and some meteorites also contain significant amounts of phyllosilicates \citep{Brearley1989, Busek1993}. 

CAIs are thought to be condensation products of the gas near the star, which had a composition that is similar to that of the gas in the planet-forming regions \citep{Bekaert2021}. The analysis of short- and long-lived isotopes, together with petrologic studies of these chondritic components, suggests that the CAIs were the first solid matter; therefore, their age can be equated with the age of the cosmochemical origin of the Solar System \citep{Lodders2005,Chaussidon2015,Bermingham2020,Connelly2012}.  Four CAIs have been dated by using the Pb-Pb chronometer, which yields a Solar System age of 4567.30 $\pm$ 0.16 Myrs \citep{Amelin2010,Connelly2012}.  Recently, \citet{Piralla2023} re-evaluated the age of the CAIs to 4568.7~Myrs through unified Al-Mg and Pb-Pb chronology.  

Chondrules are major constituents of most chondrites.  They are about millimetre-sized, round, and silicate-rich grains that are assumed to have been entirely or partly molten at some point in their history in the solar nebula.  According to \cite{Scott2005}, they crystallised in minutes to hours at temperatures of $1300-1800$\,K, prior to accretion into their parent bodies.  Chondrules can be separated in two types based on the Mg-number that is controlled by the oxygen fugacity of the chondrule-forming environment in the disc, where type I chondrules formed under more reducing conditions than type II chondrules.  A comprehensive list of minerals that were discovered in chondrules along with all their chemical variations is given in the review of \citet{Brearley1998}. 

All internal rearrangement and phase-change processes within the chondrules require high temperatures, in particular the melting.  However, refractory materials such as quartz and corundum require pressures over 8\,bar and 7000\,bar, respectively, to have a liquid phase in a solar composition gas, see Appendix\,\ref{sec:phasedia}.  At lower pressures, there is only sublimation and resublimation, but no melting.  Since such large gas pressures are not available during disc evolution, we suggest to interpret the observations of ``molten'' materials by annealing.  A true melting of refractory materials seems to be possible only inside of internally heated, larger gravitationally bound bodies such as planetesimals.

The chemical analysis of chondrules can be used for an estimation of the lifetime of the solar nebula. It is expected that most chondrules formed between $\approx 1-3$ Myr after the origin of the CAIs, in a period when the proto-Sun accreted more slowly \citep{Pape2019}.  \citet{Bollard2017} analysed the 22 youngest chondrules and determined their ages from the measurements of Pb isotopic ratios.  Bollard et al.\ concluded that the production and melting of chondrules began contemporaneously with CAI condensation, and continued for about 4 Myrs. This lifetime estimate is in good agreement with observations of protoplanetary discs, which show lifetimes of about $1-10$\,Myrs \citep{Montmerle2006,Mamajek2009}.


Isotopic data of chondritic components in the CAIs are an excellent tool for understanding the formation of the planetary building blocks as well as their mixing and transport in the disc. It is expected that CAIs formed in low-pressure environments where the temperatures are larger than about 1300\,K in an $^{16}$O-rich reducing environment \citep{Grossman1974,Krot2018,MacPherson2005}. 
Recently, \citet{Nakashima2023} analysed CAIs that were discovered in returned samples of the Japanese mission Ryugu (Hayabusa 2). It was found that three chondrule-like objects with a size of $<\!30\,\mu$m are dominated by Mg-rich olivine, and are enriched in $^{16}$O compared to $^{17}$O by $\sim\!(3-23)\permil$, which is in agreement with what has been proposed as early generations of chondrules \citep{Nakashima2023}. According to \citet{Nakashima2023} the $^{16}$O-enriched objects are most likely melted amoeboid olivine aggregates that escaped from the incorporation into the $^{16}$O-poor precursor dust of the chondrules. These authors further suggest from their analysis of the Ryugu samples that small objects experienced radial transport in the inner solar nebula to the location of the formation of the Ryugu parent body, which was farther from the Sun and scarce in chondrules. Moreover, \citet{Nakashima2023} suggest that the transported objects have been most likely destroyed during aqueous alteration in the Ryugu parent body. 
Further results from the analysis of these chondritic components include nucleosynthetic Ti isotope variations and their combination within different bulk chondrites, as well as oxygen isotope ratios.

Amoeboid olivine aggregates (AOAs) are another class of refractory objects that are related to CAIs, however, they escaped extensive melting and thus, preserved their internal structure since their formation in the pristine solar nebula.  Together with chondrules and matrix, CAIs and AOAs are the main refractory components of chondritic meteorites that apparently formed under different conditions and in different regions of the disc and therefore comprise isotopically different materials \citep{Krot2019,Amelin2010,Desch2022,Oberg2023}. 



Although there has been significant progress over recent years regarding the emergence and implications of CAIs, there are still many unanswered questions and long-standing problems related to the physical, chemical, and environmental conditions that lead to the formation of CAIs and their subsequent storage and transport in the disc \citep{MacPherson2005,Krot2015,Che2021}. 
To better understand these processes, 
we have developed a model that combines disc evolution with 2D radiative transfer, equilibrium condensation and new dust opacity calculations, see Sect.~2.
In Sects.~3 we apply this model to the early solar nebula and present the results. In Sect.~4 we discuss the implications of our findings concerning the CAI formation. Section 5 concludes our work.


\section{The model}
\begin{figure}[!b]
  \vspace*{-1mm}
  \centering
  \includegraphics[width=73mm,trim=0 0 0 0,clip]{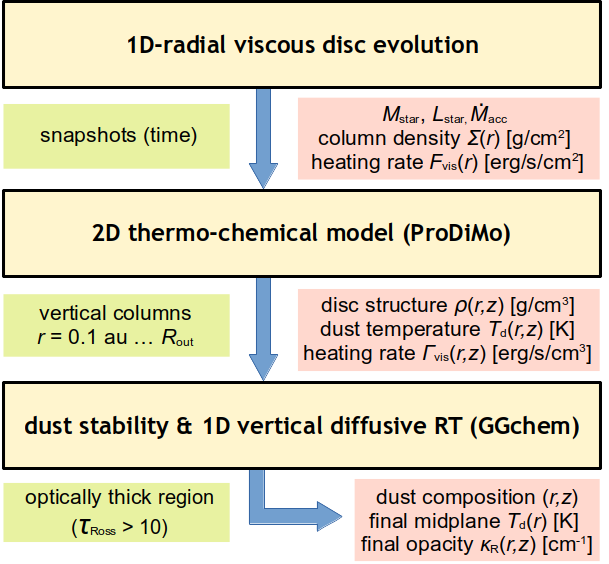} \\[-2mm]
  \caption{The three modelling stages, passed quantities, and results.}
  \label{fig:ModelStages}
\end{figure}

Our model for the early evolution of the solar system is generated in three modelling stages, see Fig.\,\ref{fig:ModelStages}.
First, we run a one-dimensional disc evolution model developed by \cite{Drazkowska2018,Drazkowska2023b} for an initial cloud mass of $1.15\,M_\odot$, which results in a stellar mass of about $1\,M_{\odot}$ after 3\,Myrs, see Sect.~\ref{sec:model1}.
Second, selecting a sequence of ages, we import the disc structures from stage~1 into the 2D radiation thermo-chemical disc model ProDiMo \citep{Woitke2009,Woitke2016,Woitke2024} and run detailed radiative transfer calculations, see Sect.~\ref{sec:model2}.
And third, we import the 2D disc structures from stage~2 into the chemical and phase equilibrium code GGchem \citep{Woitke2018} to investigate the thermal stability of the refractory dust and the ices.
In this last modelling stage, we recompute the solid opacities and modify the temperature structure in the inner, viscously heated parts of the disc close to the midplane, until the opacities are consistent with diffusive radiative transfer.  
This reveals the thermostat regulation mechanism discovered by \cite{Min2011}, where the sublimation of the main silicates leads to a loss of opacity, which determines how efficiently the viscous heat can be transported away, and thereby regulates the midplane temperatures.
In the following sections, we explain the three modelling stages in detail.
A discussion of the shortcomings of our modelling approach can be found in Sect.~\ref{sec:shorts}. 

\begin{table*}
  \caption{Disc model parameters for the early Solar Nebula}
  \label{tab:star}
  \vspace*{-3mm}\hspace*{-1.5mm}
  \resizebox{17cm}{!}{
  \begin{tabular}{cccccccccccccc}
    \hline
    &&&&&&\\[-2.2ex]
    & age & \!\![Myr]\!\! & 0.001 & 0.003 & 0.01 & 0.03 & 0.1 & 0.3 & 1 & 3 \\
    &&&&&&\\[-2.2ex]
    \hline
    &&&&&&\\[-2.2ex]
    stellar mass           & $M_{\star}$ & $[M_\odot]$
     &    0.233 &    0.236 &    0.246 &    0.275 &    0.365 &   
          0.574 &    0.887 &    0.942 \\
    accretion rate$^{(0)}$ & $\Mdot$  & \!\!\!\![$M_\odot$/yr]\!\!\!\!
     & 9.87(-7) & 9.95(-7) & 1.01(-6) & 1.03(-6) & 1.04(-6) & 
       8.31(-7) & 7.22(-8) & \!\!1.54(-8)\!\! \\
    effective temp.$^{(1)}$ & $T_{\star}$ & [K]
     &     3140 &     3160 &     3190 &     3290 &     3520 &
           3860 &     4200 &     4210 \\
    \!\!stellar luminosity$^{(1)}$\!\! & $L_{\star}$ & $[L_\odot]$
     &     2.84 &     2.68 &     2.29 &     1.83 &     2.25 &
           3.70 &     1.95 &     0.91 \\
    accretion luminosity   & $L_{\rm acc}$ & $[L_\odot]$
     &     0.50 &     0.54 &     0.64 &     0.89 &     1.26 & 
           1.52 &     0.34 &     0.11 \\
    disc luminosity        & $L_{\rm vis}$ & $[L_\odot]$
     &    0.094 &    0.097 &    0.110 &    0.142 &    0.223 &  
          0.293 &    0.048 &    0.011 \\  
    &&&&&&\\[-2.2ex]
    \hline
    &&&&&&\\[-2.2ex]
    disc mass              & $M_{\rm disc}$ & $[M_\odot]$
     &  0.00463 &  0.00495 &  0.00611 &   0.0102 &   0.0311 & 
          0.139 &    0.262 &    0.208 \\
    inner disc radius$^{\,(2)}$ & $R_{\rm in}$ & [au]
     &    0.131 &    0.128 &    0.122 &    0.118 &   0.134 &
          0.164 &    0.108 &   0.0724 \\
    outer disc radius$^{\,(3)}$ & $R_{\rm out}$ & [au]
     &      17  &       18 &       20 &       25 &      42 &
            82  &      260 &      630 \\
    outer disc radius$^{\,(4)}$ & \!\!\!$R_{\rm out,obs}$\!\!\! & [au]
     &      4.8 &      5.0 &      5.5 &      6.3 &     9.9 &
             25 &       81 &      130 \\
    &&&&&&\\[-2.2ex]
    \hline
  \end{tabular}}\\[1mm]
  \small $^{(0)}$: notation $a(-b)$ means $a\times 10^{-b}$\\ 
  \small $^{(1)}$: from the evolutionary tracks of \cite{Siess2000} for given age and $M_\star$\\
  \small $^{(2)}$: using $(L_\star\!+\!L_{\rm acc})\propto R_{\rm in}^2$, see text\\
  \small $^{(3)}$: end of the ProDiMo model domain, corresponding to a half column density of hydrogen nuclei of $N_{\rm\langle H\rangle}\!=\!10^{\,20}\rm\,cm^{-2}$\\
  \small $^{(4)}$: radius that contains 90\% of the 1.3\,mm continuum-flux from the disc according to the ProDiMo model
\end{table*}

\subsection{The evolutionary disc model}
\label{sec:model1}

We use a hydrodynamic disc evolution model similar to the one described in \citet{Dullemond2006a, Dullemond2006b, Drazkowska2018,Drazkowska2023b}. Our model considers the formation of a star and circumstellar disc from a single parent molecular cloud core with a mass of $1.15\,M_\odot$, constant temperature of 10\,K, and rotation rate of $7\times10^{-15}$~s$^{-1}$. The collapse proceeds from inside-out.  We assume that angular momentum is conserved during the infall and thus the material from every shell of the parent cloud falls onto the disc inside of the centrifugal radius.  The centrifugal radius moves outward with time, which strongly depends on the rotation rate of the parent cloud.  The lower the initial cloud rotation rate, the smaller the centrifugal radius, and thus the material falls on the hot part of the disc for a longer time. The way the material is distributed inside of the centrifugal radius is described in \citet{Hueso2005}. We assume that all material falling inside of the disc's inner radius, set here to 0.1\,au, is added directly to the central star, thus the star is fed both directly from the molecular cloud and through the disc accretion. The inside-out infall leads to an initially compact and hot disc, which reaches its maximum mass of $0.3\,M_\odot$ after about 0.6\,Myrs.  

The disc surface density evolution is calculated by taking into account the mass infall from the parent cloud and viscous evolution which we parameterise using the standard $\alpha$-formalism and setting the global value to $\alpha=10^{-3}$ \citep{1973A&A....24..337S}. During the infall, the compact disc vigorously expands and the outward-flowing gas drags millimetre-sized solid particles with it.  Such disc models have been proven to produce the right conditions for outward transport of the CAIs necessary to explain their incorporation in meteorites formed in the outer parts of the solar system \citep{Cuzzi2003, Ciesla2007, Jacquet2019, Jongejan2023}, however, the formation of CAIs was not studied consistently in these models, instead these were added afterwards.


\subsection{The ProDiMo disc model}
\label{sec:model2}

The \underline{Pro}toplanetary \underline{Di}sc \underline{Mo}del (ProDiMo) developed by \cite{Woitke2009, Woitke2016, Woitke2024} is a state-of-the-art code that combines disc chemistry with radiative transfer, heating/cooling and high-energy physics.  The considered chemical processes include 2-body and 3-body gas-phase chemistry \citep{Kamp2017}, X-ray ionisation \citep{Aresu2011}, FUV and EUV photo-processes \citep{Rab2017}, and ice chemistry. The ProDiMo models in this paper are based on the radial disc structures imported from the evolutionary disc model (Sect.~\ref{sec:model1}) for a sequence of eight consecutive ages $0.001, 0.003, 0.01, 0.03, 0.1, 0.3, 1,$ and 3\,Myrs, where time zero is the beginning of the cloud collapse. 
For each age we use the following data as input: the stellar mass $M_\star$, the stellar mass accretion rate $\dot{M}$, the disc surface density profile $\Sigma(r)\rm\,[g/cm^2]$, and the vertically integrated viscous heating rate $F_{\rm vis}(r)\rm\,[erg/s/cm^2]$.   

\subsubsection{Stellar setup}

The stellar mass $M_\star$ and the mass accretion rate $\dot{M}_{\rm acc}$ are taken from the evolutionary disc model (Sect.~\ref{sec:model1}). 
The stellar luminosity $L_\star$ and the effective temperature of the star $T_\star$ are interpolated between the values given in the pre-main sequence evolutionary tracks of \cite{Siess2000} for given age and $M_\star$, see Table~\ref{tab:star}.
We note that there is considerable uncertainty in the stellar luminosity at the earliest stages since the protostar is deeply embedded within the natal cloud and thus, not directly observable \citep{White+07}.
The derived values of $L_\star$ and $T_\star$ depend on the particular pre-main sequence evolutionary model selected to calculate the tracks.
When comparing against the tracks of \cite[][]{YB08}, we find a good agreement concerning $T_\star$, but the stellar luminosities differ notably before 0.3 Myr, with luminosities being almost one order of magnitude lower at the earliest evolutionary stages. However, these disagreements diminish with age and differ only about 10\% at later times. 

The stellar radius $R_\star$ is calculated from the Stefan–Boltz\-mann law $L_\star\!=\!4\pi\,R_\star^2\,\sigma T_\star^4$, where $L_\star$ represents the energy liberated by the contraction of the protostar and the beginning nuclear fusion.
The accretion luminosity $L_{\rm acc}$ represents the energy liberated by the gas falling onto the star, from the inner disc radius $R_{\rm in}$ to the stellar radius $R_\star$.  It is calculated as
\begin{equation}
  L_{\rm acc} = \frac{G M_\star\dot{M}_{\rm acc}}{2\,R_\star}
              \left(1-\frac{R_\star}{R_{\rm in}}\right) \ .
  \label{eq:Lacc}
\end{equation}
The inner radius of the disc $R_{\rm in}$ is set to where the total stellar irradiation causes the dust to sublimate, considering a dust temperature of about 1500\,K 
\begin{equation}
  R_{\rm in} = \big(0.1\,{\rm au}\big)
             \sqrt{\frac{L_\star+L_{\rm acc}}{1.95\,L_\odot}} \ .
  \label{eq:Rin}
\end{equation}
The value of 0.1\,au for a total luminosity of 1.95\,$L_\odot$ comes from previous experience with the ProDiMo disc models.  Equations (\ref{eq:Lacc}) and (\ref{eq:Rin}) are solved iteratively until converged. Table~\ref{tab:star} summarises the stellar setup parameters for the ProDiMo models.  For comparison, the table also contains the viscous disc luminosity 
\begin{equation}
    L_{\rm vis} = \int_{R_{\rm in}}^{R_{\rm out}} 
                  F_{\rm vis}(r)\,2\pi r\,dr \ ,
\end{equation}
where $F_{\rm vis}$ is the viscous heating rate per surface area, and $R_{\rm out}$ is the outer disc radius. Our results indicate that $L_{\rm vis}\!<\!0.1 L_\star$ at any evolutionary state considered here. 

The stellar spectra required for the ProDiMo radiative transfer calculations are compiled from two components: (i) the photospheric spectrum, representing $L_\star$ and $T_\star$, and (ii) a blackbody of $T_{\rm acc}\!=\!10000\,$K, representing the accretion luminosity $L_{\rm acc}$ produced by the accretion shock.  The total spectral intensities emitted from the star's surface are calculated as
\begin{equation}
  I_\nu^\star ~=~ 
  I_\nu^{\rm phot}\big(T_\star,\log\,g\big) 
  ~+~ \frac{L_{\rm acc}}{4\pi\,R_\star^2\,\sigma T_{\rm acc}^4} \,B_\nu(T_{\rm acc}) \ .
\end{equation}
The photospheric intensities $I_\nu^{\rm phot}$ are interpolated from a grid of PHOENIX stellar atmosphere models by \cite{Brott2005}, where the surface gravity is given by $g\!=\!GM_\star/R_\star^2$.

\subsubsection{Disc structure, dust settling, and radiative transfer}

Based on the input gas surface density structure $\Sigma(r)$, ProDiMo computes an axisymmetric 2D density structure $\nH(r,z)$, where $\nH\rm\,[cm^{-3}]$ is the hydrogen nuclei density, in vertical hydrostatic equilibrium as follows.  For each column $r$, starting with a guess of the temperature structure $T(r,z)$, ProDiMo numerically integrates the equation of hydrostatic equilibrium upwards \citep{Woitke2009}
\begin{equation}
  \frac{1}{\rho}\frac{dp}{dz} = -\,\frac{z\;GM_\star}{\big(r^2+z^2\big)^{3/2}} \ ,
  \label{eq:hydrostat} 
\end{equation}
using the ``simplified hydrostatic model'' \citep{Woitke2016}, where the isothermal sound speed $c_T = \sqrt{p/\rho} = \sqrt{kT/(2.3\,{\rm amu})}$ is evaluated with $T\!=\!\Td$ and a constant (\ce{H2}-rich) mean molecular weight. The resulting density structure $\rho(r,z)$ is then integrated vertically and normalised to match the given $\Sigma(r)$.
The differences to a ``full hydrostatic model'', where the gas temperature $\Tg$ and the proper mean molecular weight are used to calculate the gas pressure, are visualised in Fig.~11 of \cite{Woitke2016}. Since we focus on the dust evolution in the midplane in this paper, where $\Tg\!=\!\Td$ is valid, this approximation has little effect on the results.

Next, ProDiMo sets up the settled dust structure and calculates the frequency-dependent dust opacities as explained in \cite{Woitke2016}.  We start with an unsettled size distribution function with a powerlaw $f(a)\!\propto\!a^{-\apow}$ between a minimum ($\amin$) and a maximum dust size ($\amax$). Using 100 dust size bins, the dust in each bin is then vertically redistributed (settled) in each column according to the given gas density structure, the turbulent mixing parameter $\aset$, and the settling description of \cite{Riols2018}, see \cite{Woitke2024} for details.  We used the following parameter values in this paper: $\amin\!=\!0.05\rm\,\mu$m, $\amax\!=\!3\rm\,$mm, $\apow\!=\!3.5$, dust-to-gas mass ratio\,$=\!0.004$, and $\aset\!=\!10^{-3}$.  The dust opacities are then calculated by Mie theory, assuming a constant mixture of 60\% silicates, 15\% amorphous carbon and 25\% porosity, using the settled dust size distribution function $f(a,r,z)$ at each point of the model.

To compute $\Td(r,z)$, ProDiMo uses its in-built frequency-dependent ray-based 2D radiative transfer module, which has been thoroughly tested against the leading Monte-Carlo radiative transfer programs by \cite{Pinte2009}, see also \cite{Woitke2016} and Appendix~A in \citet{Oberg2022}. 
In the optically thick and viscously heated midplane, most relevant for this paper, the conservation of the frequency-integrated radiative flux can be formulated as
\begin{equation}
  \div \vec{F} ~=~ \Gamma \ ,
  \label{eq:divF}
\end{equation}
where $\Gamma$ is the local net heating rate converting non-radiative to radiative energy.
The input model (Sect.~\ref{sec:model1}) provides the viscous heating rate per unit area $F_{\rm vis}\,\rm[erg/cm^2/s]$. To calculate the heating rate per volume $\Gamma$, we assume that the heating occurs proportional to the gas density $\rho$ as
\begin{equation}
  \Gamma(r,z) ~=~ \frac{1}{2} F_{\rm vis}(r) \frac{\rho(r,z)}
                  {\int_{\,0}^\infty \rho(r,z)\,dz}\ ,
  \label{eq:Fvis}
\end{equation}
where the factor $1/2$ is because of the mirror symmetry at $z\!=\!0$. In the diffusion approximation, the bolometric radiation flux 
\begin{equation}
  \vec{F}(r,z) = -\frac{4\,\pi}{3\kRoss(r,z)}\,\grad\,J(r,z)
  \label{eq:diffApprox}
\end{equation}
is given by the gradient of the bolometric mean intensity $J(r,z)\!=\!\sigma\,T^4/\pi$, where $\sigma$ is the Stefan-Boltzmann constant. The Rosseland-mean opacity is defined as
\begin{equation}
  \kRoss(r,z) =
    \int \frac{dB_\nu(T)}{dT}\,d\nu \;\;\bigg/\;
    \int \frac{1}{\kappa^{\rm ext}_\nu(r,z)} \frac{dB_\nu(T)}{dT}\,d\nu
  \label{eq:kRoss}  
\end{equation}
where $\kappa^{\rm ext}_\nu(r,z)$ is the dust extinction coefficient at position $(r,z)$ in the disc, $T\!=\!T_{\rm dust}(r,z)$ is the temperature, and $B_\nu(T)$ is the Planck function. ProDiMo solves these diffusion equations in 2D for given opacity structure in the optically thick core of the midplane, using the temperature results from the proper 2D frequency-dependent ray-based radiative transfer as upper boundary condition. We assume that the vertical component of the flux is zero in the midplane, because of symmetry, and use this condition as lower boundary condition, see Appendix~A in \citet{Oberg2022}.

Once the 2D disc radiative transfer problem is solved, and the dust temperature structure $\Td(r,z)$ is determined, ProDiMo calculates the chemical composition in time-independent mode and the gas temperature structure $\Tg(r,z)$ in heating/cooling balance.
These results are not relevant to the problem of dust stability in the midplane, and will hence not be discussed any further in this paper.
However, $\Td(r,z)$ is required for the computations of the vertical hydrostatic equilibrium (Eq.~\ref{eq:hydrostat}), and therefore, ProDiMo performs global iterations between radiative transfer and hydrostatic equilibrium, which converges after about 10 iterations.

\subsection{The GGchem phase equilibrium model}
\label{sec:model3}

The basic idea for this paper is to consider the process of refractory dust sublimation as post-processing of the ProDiMo disc models, using GGchem \citep{Woitke2018}. GGchem is a publicly available thermo-chemical equilibrium code to calculate the concentrations of all neutral and single ionised atoms, electrons, molecules, molecular ions, and condensates.  GGchem is based on the principle of minimisation of the total Gibbs free energy, including condensation, and applicable in a wide temperature range between 100\,K and 6000\,K. For this paper, we use the code down to 50\,K; for temperatures below 50\,K we use $T\!=\!50K$.

The sole purpose of GGchem here is to predict the amount and volume composition of all solids, for given gas density $\nH$, temperature $T$, and total element abundances prior to condensation, where we use the solar abundances of \cite{Asplund2009}. The GGchem setup in this paper is the same as in \cite{Woitke2018}. We select 26 elements (H, He, Li, C, N, O, F, Na, Mg, Al, Si, P, S, Cl, K, Ca, Ti, V, Cr, Mn, Fe, Ni, Cu, Zn, Zr, and W) and include condensates, ions and charged molecules, which leads to a selection of 595 gaseous chemical species and 248 condensed species.

For this paper, we have extended GGchem to calculate the opacity of the resulting mixture of solid materials as explained in Sect.~\ref{sec:opacities}.  However, that opacity structure changes the diffusive radiative transfer in the optically thick core of the disc, and therefore, dust stability and radiative transfer become a coupled problem.  The following section describes how we solve this issue using a simplified vertical diffusive radiative transfer, which results in a thermostat regulation mechanism, which changes $T(r,z)$ in the optically thick midplane, and only there.

\subsubsection{The thermostat regulation mechanism}
\label{sec:thermostat}

When the midplane is very optically thick and heated by frictional processes, the radiative flux is almost perfectly directed upwards in the vertical direction.  In this case, we can simplify Eqs.\,(\ref{eq:divF}) and (\ref{eq:diffApprox}) as
\begin{equation}
  F(z) ~=~ \int_0^z \Gamma(z')\,dz'
       ~=~ -\frac{4\,\sigma}{3\kRoss(z)}\,\frac{dT^4}{dz} ,
  \label{eq:1Dflux}
\end{equation}
and hence the vertical flux $F(z)$ is known because we know $\Gamma(z)$.
Considering a vertical grid $\{z_k\,|\,k\!=\!0,...\,,K\}$, taken over from the ProDiMo disc model, Eq.\,(\ref{eq:1Dflux}) is written numerically as
\begin{equation}
  F(z_{k-\half})    
  ~=~ -\frac{4\,\sigma}{3\sqrt{\kRoss(z_{k-1})\,\kRoss(z_k)}}
      \,\frac{T(z_k)^4-T(z_{k-1})^4}{z_k-z_{k-1}} \ ,
  \label{eq:ToSolve}
\end{equation}
where $z_{k-\half} = (z_{k-1}+z_k)/2$. Equation\,(\ref{eq:ToSolve}) can be solved downwards ($k\!\to\!k-1$) to obtain the consistent temperature and opacity structure. For each disc column at radius $r$, we start the integration at a height $z_K$ over the midplane, where the vertical and horizontal Rosseland optical depth are both $>\!10$, to make sure we are safely in the diffusive regime.  We then read off $T(z_K)$ at that point as upper boundary condition, and the density structure $\nH(z)$ below ($z\!\leq\!z_K$) from the ProDiMo disc model. Then, we
\begin{itemize}
\setlength{\parskip}{0.5pt}
\item[1.] guess the temperature $T(z_{k-1})$,
\item[2.] compute the amount of dust and its material composition for given total element abundances, temperature $T(z_{k-1})$ and hydrogen nuclei particle density $\nH(z_{k-1})$ with GGchem,
\item[3.] calculate the dust size distribution function $f(a)$ at $z_{k-1}$ (Sect.~\ref{sec:sizedist}),
\item[4.] calculate the frequency-dependent dust extinction coefficient $\kappa^{\rm ext}_\nu(z_{k-1})$ according to Eq.\,(\ref{eq:kappa}) by effective medium and Mie theory (Sect.~\ref{sec:opacities}), 
\item[5.] calculate the Rosseland mean opacity $\kRoss(z_{k-1})$ according to Eqs.\,(\ref{eq:kRoss}) and (\ref{eq:kRossTot}),
\item[6.] calculate $T(z_{k-1})$ according to Eq.\,(\ref{eq:ToSolve}),
\item[7.] go back to step~2 if not yet converged.
\end{itemize}  
This way, we have implemented a stable thermostat regulation mechanism as discovered by \cite{Min2011}.  If the temperature $T(z_{k-1})$ becomes too high, the dust starts to sublimate, which lowers the Rosseland opacity $\kRoss$, and hence limits the downward temperature increase. In all cases where both the viscous heating rate $F_{\rm vis}$ and the vertical optical depths are large enough, the solution will always lead to silicate sublimation in the deeper layers, but rarely to complete dust evaporation, as some of the most stable condensates will remain: the Al-Ca-Ti oxides.

\subsubsection{Dust size distribution}
\label{sec:sizedist}

For the dust size distribution function $f(a)$, we consider a reference dust size distribution function $f_{\rm ref}(a)\!\propto a^{-3.5}$ (in units $\rm cm^{-1}/$H-nucleus) between 0.05 and 3000\,$\mu$m, which is normalised to a dust/gas mass ratio of 0.004, assuming an interstellar dust material density of $\rm 2\,g/cm^3$. The value 0.004 results from complete condensation of a solar composition gas \citep{Asplund2009} before ice formation, see \citet{Woitke2018} for details.  The reference dust size distribution function is the same as the unsettled dust size distribution function in the ProDiMo disc model. The total reference volume of condensed species is
\begin{align}
  V_{\rm ref} ~=&~ \int_{0.05\rm\,\mu m}^{3\rm mm}
  \frac{4\pi}{3}a^3\,f_{\rm ref}(a)\,da \\
            ~=&~~ 4.53\times10^{-27}\,\mbox{cm$^3$/H-nucleus} \ . \nonumber
\end{align}
In the phase equilibrium simulations, GGchem provides a different total volume of condensed species $V$ per H-nucleus, depending on which condensed materials are found to be stable, and how much of them.  If $V\!<\!V_{\rm ref}$, which happens, for example, when the silicates sublimate, we assume that all particles have shrunk by a common factor in radius
\begin{equation}
  \gamma = \left(\frac{V}{V_{\rm ref}}\right)^{1/3} \ .
  \label{eq:gamma}
\end{equation}
The numerical grid for the size distribution function $\{(a_i,f_i)\,|\,i=0,...,I\}$ is then modified as $a_i\!=\!\gamma\times a_{{\rm ref},i}$, while the number of dust particles between size gridpoints $a_i$ is assumed to remain the same. However, since $f(a)$ is the number of dust particles per size interval, we have $f_i\!=\!f_{{\rm ref},i}/\gamma$. If ice condenses on the existing surfaces, leading to $V\!>\!V_{\rm ref}$, we assume that all grains increase in size accordingly, which leads to increased dust opacities per hydrogen nucleus according to Eq.\,(\ref{eq:kappa}), first and foremost because of the increased condensed volume per H-nucleus.

\subsubsection{Dust and gas opacities}
\label{sec:opacities}

The phase equilibrium model GGchem provides the volume mixing ratios $\{V_s\,|\,s\!=\!1,...,S\}$, where $s$ is an index for the stable solids, see examples in Table~\ref{tab:Vs}. We add 25\% porosity. 
Using the effective medium theory of \cite{Bruggeman1935}, we calculate the effective optical constants of the mixed material as 
\begin{equation}
  (\neff,\keff) = F(V_s,n_s,k_s)
\end{equation}
where $(n_s,k_s)$ are the real and imaginary parts of the optical constants of the pure materials $s$.  The dust extinction, absorption and scattering opacities [1/cm] are then calculated as
\begin{align}
\kappa^{\rm ext}_\nu =&~~ \nH\int_\amin^\amax f(a)
                       \pi\,a^2 Q^{\rm ext}_\nu(x,\neff,\keff)\,da
\label{eq:kappa}\\
\kappa^{\rm abs}_\nu =&~~ \nH\int_\amin^\amax f(a)
                       \pi\,a^2 Q^{\rm abs}_\nu(x,\neff,\keff)\,da \\
\kappa^{\rm sca}_\nu =&~~ \nH\int_\amin^\amax f(a)
                       \pi\,a^2 Q^{\rm sca}_\nu(x,\neff,\keff)\,da
\end{align}
where $a$ is the radius of a dust particle and $x\!=\!2\pi\,a/\lambda$ is the size parameter.  $f(a)\,$[1/cm/H-nucleus] is the size distribution function per hydrogen nucleus, see Sect.~\ref{sec:sizedist}. The extinction, absorption and scattering efficiencies are calculated according to Mie theory. We use the routine MIEX \citep{Voshchinnikov2002} in the implementation of Sebastian Wolff. The opacity computations are based on a collection of optical data by \citet{Marigo2024}, with additional data collected by Carsten Dominik and Michiel Min (\url{https://github.com/cdominik/optoo}). The complete selection of relevant materials and opacity sources is summarised in Table~\ref{tab:opac}.

\begin{figure*}
  \hspace*{-3mm}\vspace*{-3mm}
  \begin{tabular}{cccc}
  \rotatebox{90}{\hspace*{19mm}\sf 0.03\,Myr} &
    \hspace*{-4mm}
    \includegraphics[page=5,width=60.5mm,trim=30 35 55 420,clip] {Figs/out_3E4.pdf} &
    \hspace*{-6mm}
    \includegraphics[page=9,width=60.5mm,trim=29 32 52 420,clip] {Figs/out_3E4.pdf} & 
    \hspace*{-6mm}
    \includegraphics[width=60.5mm,trim=30 34 55 424,clip]{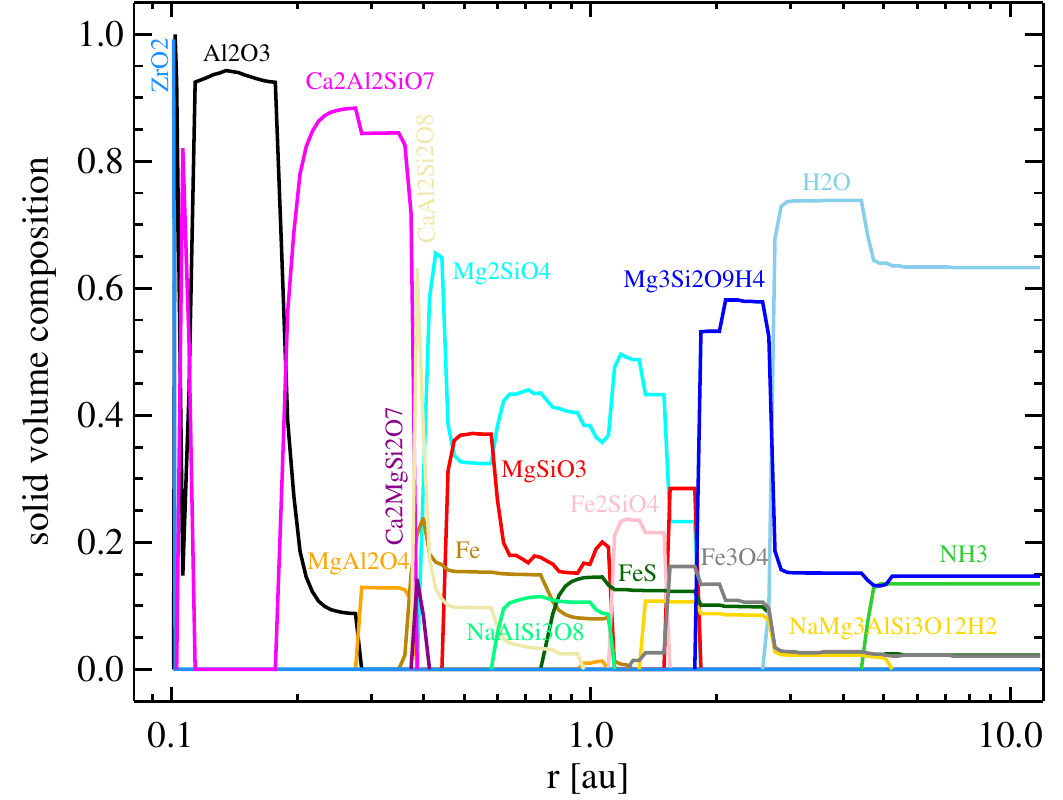} \\[-1mm]
  \rotatebox{90}{\hspace*{20mm}\sf 0.1\,Myr} &
    \hspace*{-4mm}
    \includegraphics[page=5,width=60.5mm,trim=30 35 52 420,clip] {Figs/out_1E5.pdf} & 
    \hspace*{-6mm}
    \includegraphics[page=9,width=60.5mm,trim=29 32 55 420,clip] {Figs/out_1E5.pdf} &
    \hspace*{-6mm}
    \includegraphics[width=60.5mm,trim=30 35 55 424,clip]{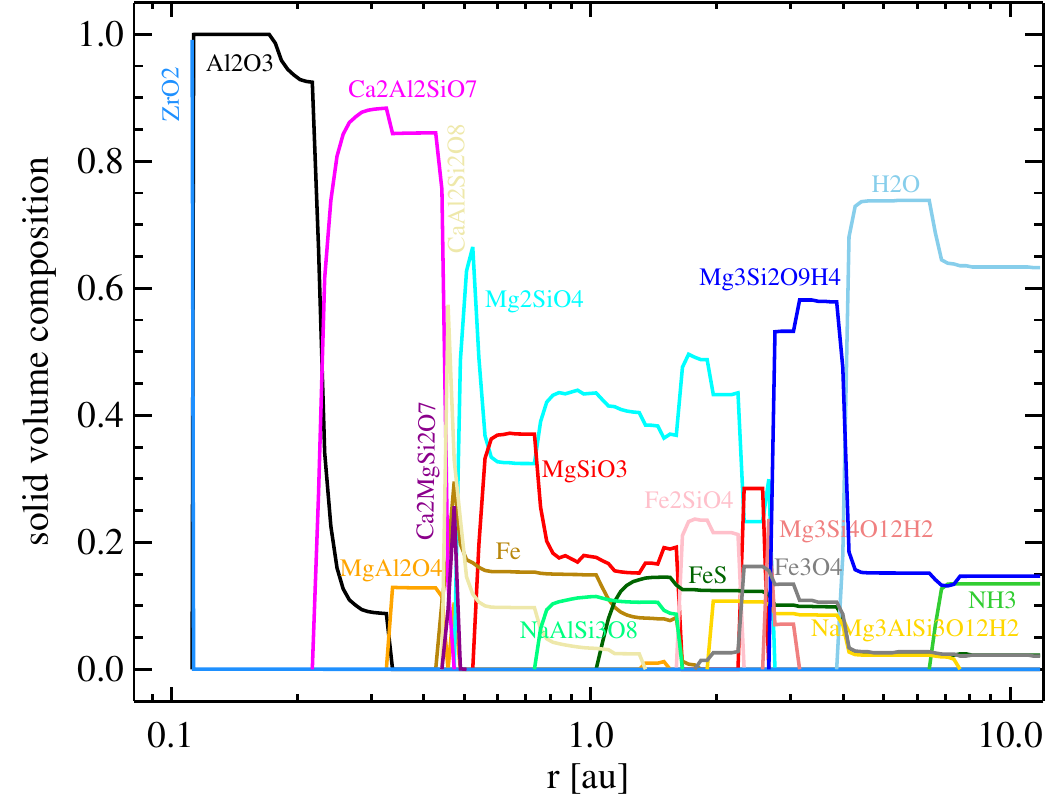} \\[-1mm]
  \rotatebox{90}{\hspace*{20mm}\sf 0.3\,Myr} &
    \hspace*{-4mm}
    \includegraphics[page=5,width=60.5mm,trim=30 35 52 420,clip] {Figs/out_3E5.pdf} & 
    \hspace*{-6mm}
    \includegraphics[page=9,width=60.5mm,trim=29 32 55 420,clip] {Figs/out_3E5.pdf} &
    \hspace*{-6mm}
    \includegraphics[width=60.5mm,trim=30 35 55 424,clip]{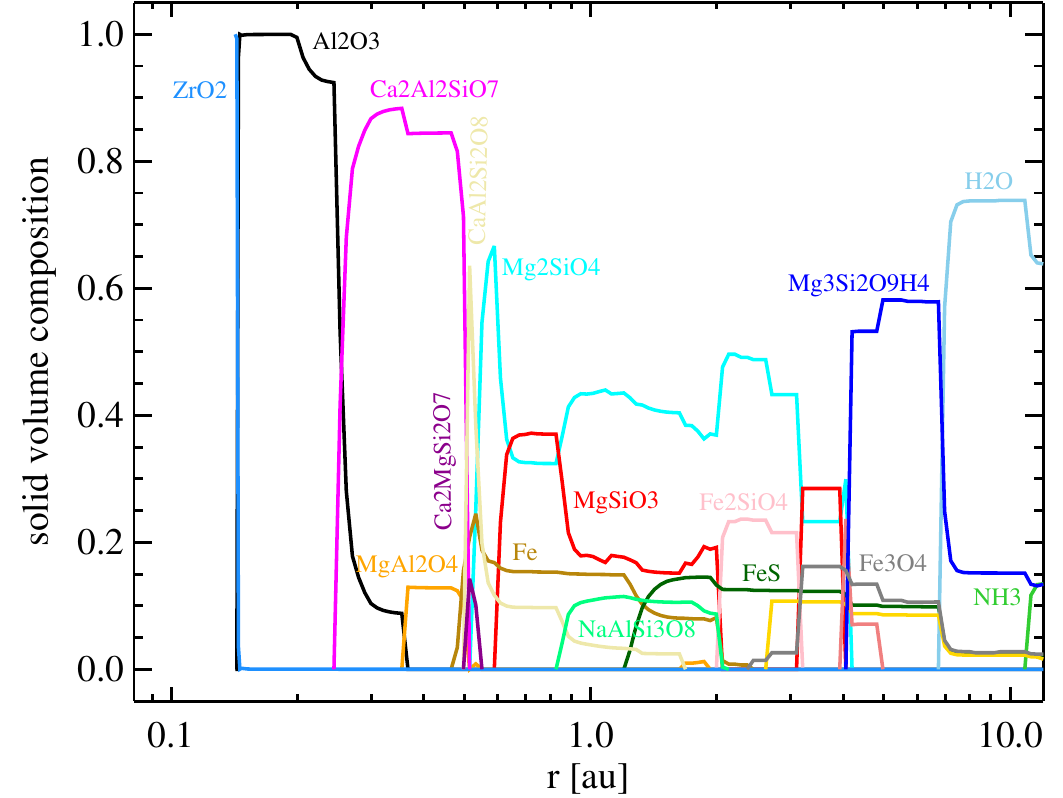} \\[-1mm]
  \rotatebox{90}{\hspace*{21mm}\sf 1\,Myr} &
    \hspace*{-4mm}
    \includegraphics[page=5,width=60.5mm,trim=30 35 52 420,clip] {Figs/out_1E6.pdf} &
    \hspace*{-6mm}
    \includegraphics[page=9,width=60.5mm,trim=29 32 55 420,clip] {Figs/out_1E6.pdf} &
    \hspace*{-6mm}
    \includegraphics[width=60.5mm,trim=30 35 55 424,clip]{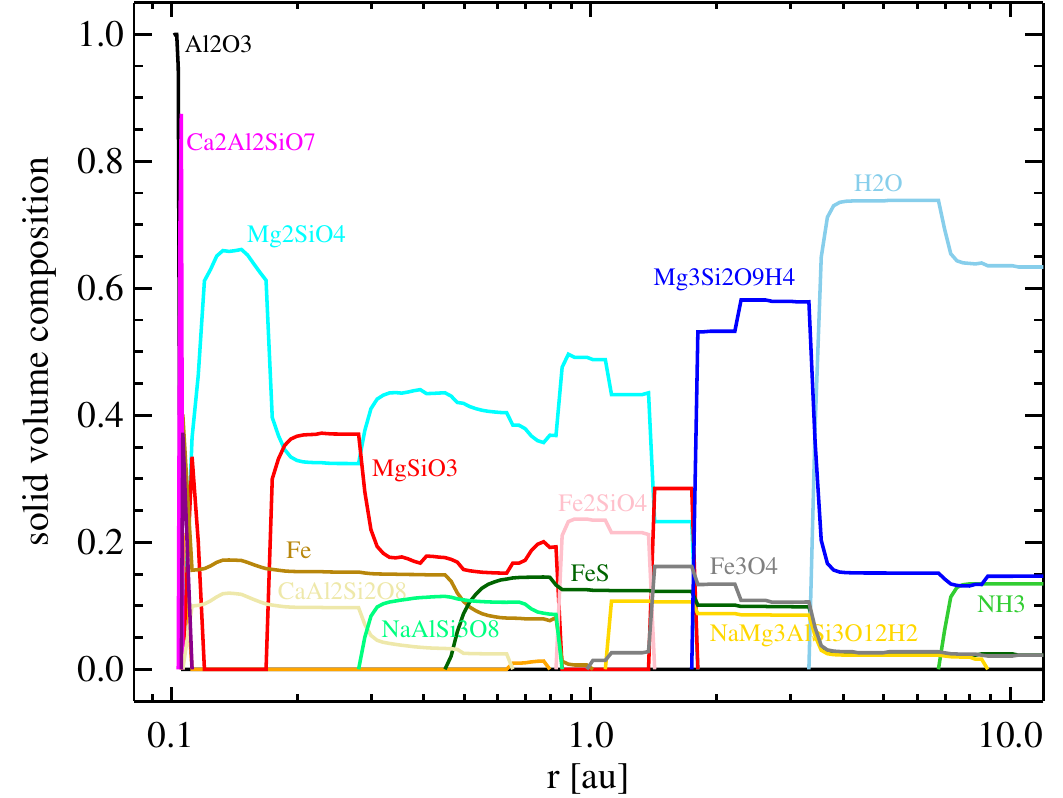} \\[-1mm]
  \rotatebox{90}{\hspace*{21mm}\sf 3\,Myr} &
    \hspace*{-4mm}
    \includegraphics[page=5,width=60.5mm,trim=30 35 52 420,clip] {Figs/out_3E6.pdf} &
    \hspace*{-6mm}
    \includegraphics[page=9,width=60.5mm,trim=29 32 55 420,clip] {Figs/out_3E6.pdf} &
    \hspace*{-6mm}
    \includegraphics[width=60.5mm,trim=30 35 55 424,clip]{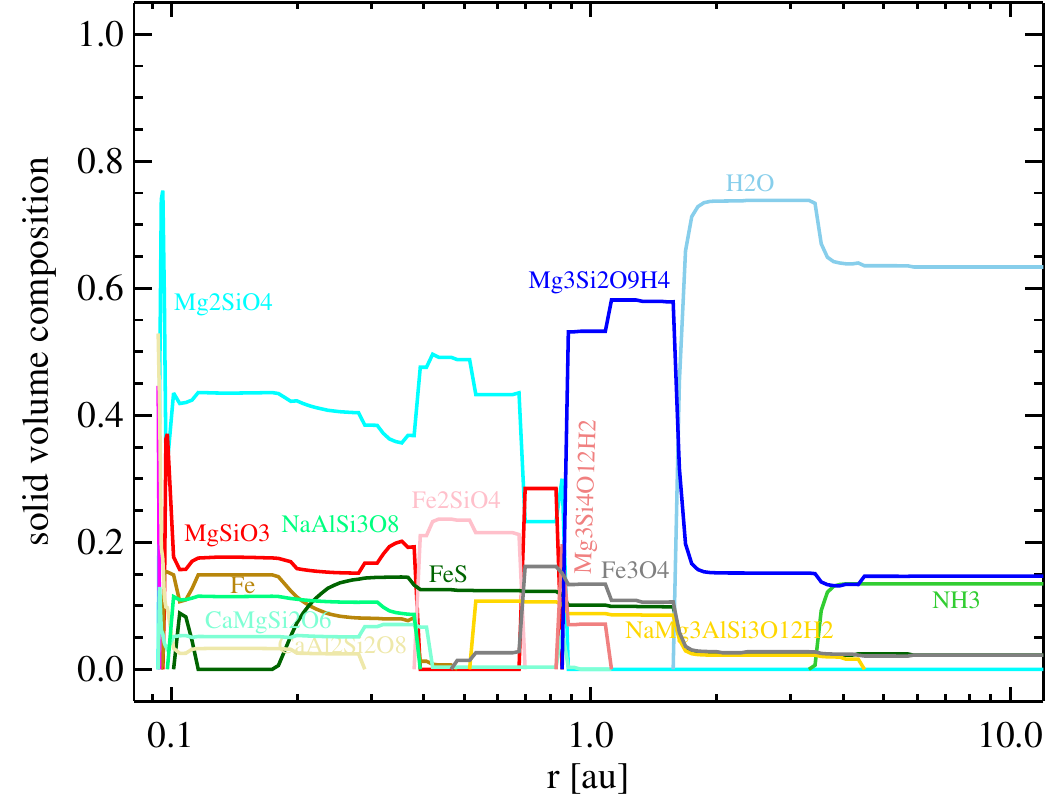} \\[1mm]
  \end{tabular}
  \caption{Evolution of gas and dust column densities ({\bf left column}) and midplane temperatures ({\bf middle column}) with ages indicated on the left. The green dots on the left mark the popular MMSN-value of $\rm 1700\,g/cm^2$ at 1\,au.  
  The dashed cyan lines are the resulting midplane temperatures from the respective ProDiMo models, and the thick magenta lines are the midplane temperatures after dust sublimation.  The dotted black line indicates the silicate stability temperature, see text.
  The {\bf right column} shows the material composition of the solid particles in the midplane.} 
  \label{fig:Temp}
  \vspace*{-2mm}
\end{figure*}

Finally, the total Rosseland opacity used in Eq.\,(\ref{eq:ToSolve}) is assumed to be given by the dust Rosseland opacity according to Eqs.\,(\ref{eq:kRoss}) and (\ref{eq:kappa}), and a gas Rosseland opacity
\begin{equation}
  \kRoss ~=~ \kRoss^{\rm dust}(\nH,V,V_s,T) ~+~ \kRoss^{\rm gas}\big(\nH,T\big) \ .
  \label{eq:kRossTot}
\end{equation}
The gas Rosseland mean opacity $\kRoss^{\rm gas}$ is interpolated from 2D-tables provided by Paola Marigo, using the same opacity data sources and techniques explained in \citet{2022ApJ...940..129M,Marigo2024}, but without the dust opacities. 
In those works we updated and expanded molecular absorption to include 80 species, predominantly utilising the recommended line lists currently accessible from the \texttt{ExoMol} \citep{EXOMOL_2012MNRAS.425...21T} and \texttt{HITRAN} \citep{HITRAN2020_GORDON2022107949} databases. Additionally, in response to a recent investigation, we revised the H$^-$ photodetachment cross section, incorporated the free-free absorption of negative ions of atoms and molecules, and refreshed the collision-induced absorption due to H$_2$/H$_2$, H$_2$/H, H$_2$/He, and H/He pairs.
To construct the Rosseland mean opacity tables, we integrated the \texttt{\AE SOPUS 2.0} code \citep{Marigo2024} with the \texttt{GGchem} code \citep{Woitke2018}. We adopted a solar mixture following \citet{Asplund2009}, with a hydrogen abundance (in mass fraction) $X=0.7374$, and metallicity $Z=0.0134$. We consider the number density of hydrogen (range: $6 \leq \log(\nH) \leq 20$ in steps of $0.1$ dex) and the temperature (range: $1.7\!\leq\!\log(T/K)\!\leq\!4.0$ in steps of $0.01$\,dex) as independent variables.

The gas opacity provides a minimum base opacity, mostly provided by water vapour line opacity, which is reached when all dust has sublimated.  The inclusion of the gas opacity means that there is an upper limit for the thermostat regulation mechanism, see Sect.~\ref{sec:regulate}.

\section{Results}
\begin{figure*}
\centering
  \begin{tabular}{cc}
    0.1\,Myr & 1\,Myr \\
    \includegraphics[ page=1,width=76mm,trim=40 43 55 420,clip]
                    {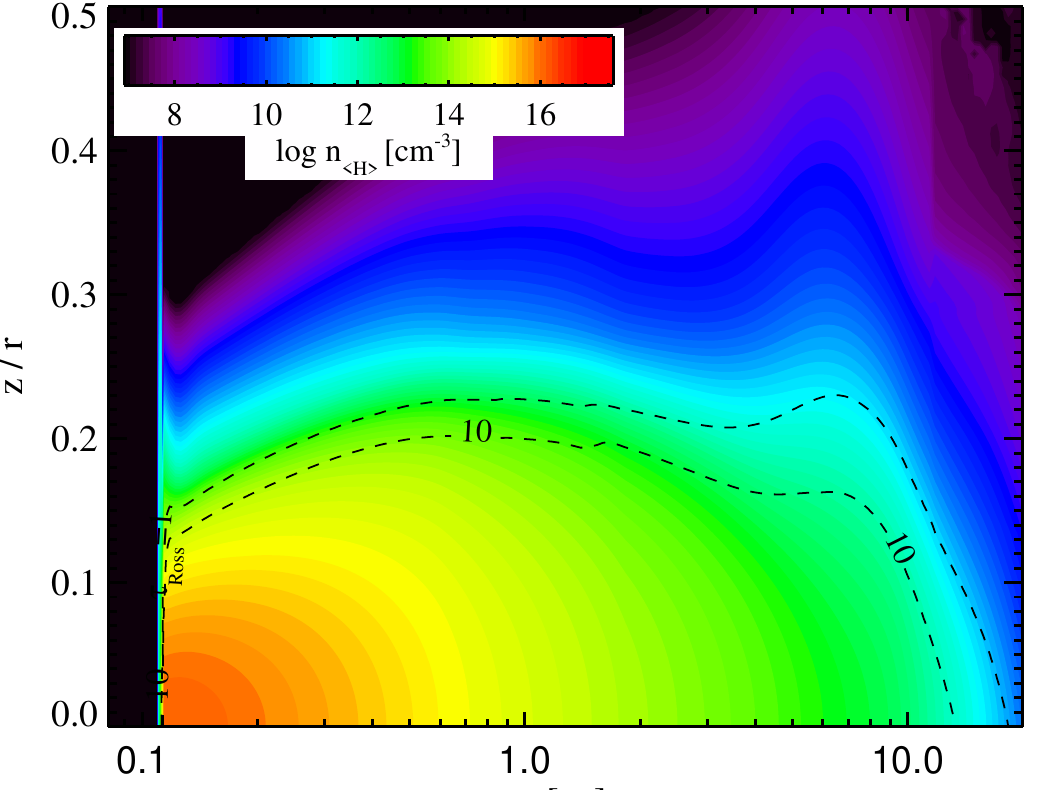} &
    \includegraphics[ page=1,width=76mm,trim=40 43 55 420,clip]
                    {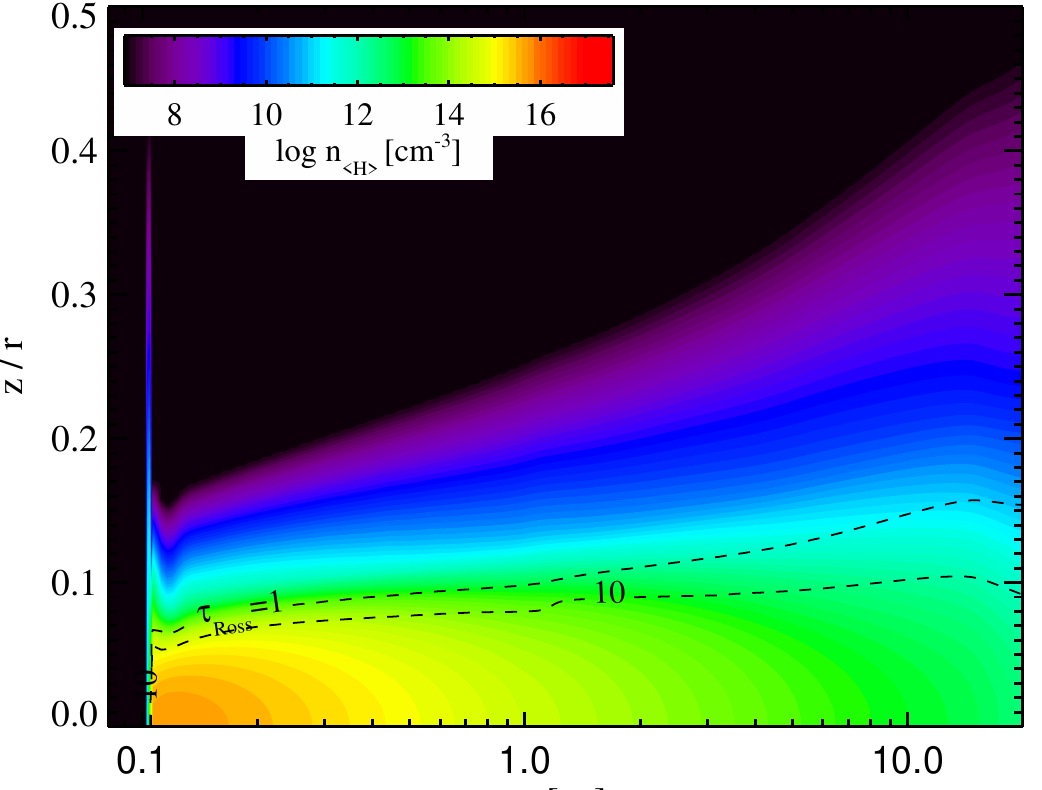} \\[-2mm]
    \includegraphics[ page=7,width=76mm,trim=40 43 55 420,clip]
                    {Figs/out_1E5.pdf} &
    \includegraphics[ page=7,width=76mm,trim=40 43 55 420,clip]
                    {Figs/out_1E6.pdf} \\[-2mm]
    \includegraphics[ page=3,width=76mm,trim=40 43 55 420,clip]
                    {Figs/out_1E5.pdf} &
    \includegraphics[ page=3,width=76mm,trim=40 43 55 420,clip]
                    {Figs/out_1E6.pdf} \\[-2mm]
    \includegraphics[width=76mm,trim=143 20 20 85,clip]
                    {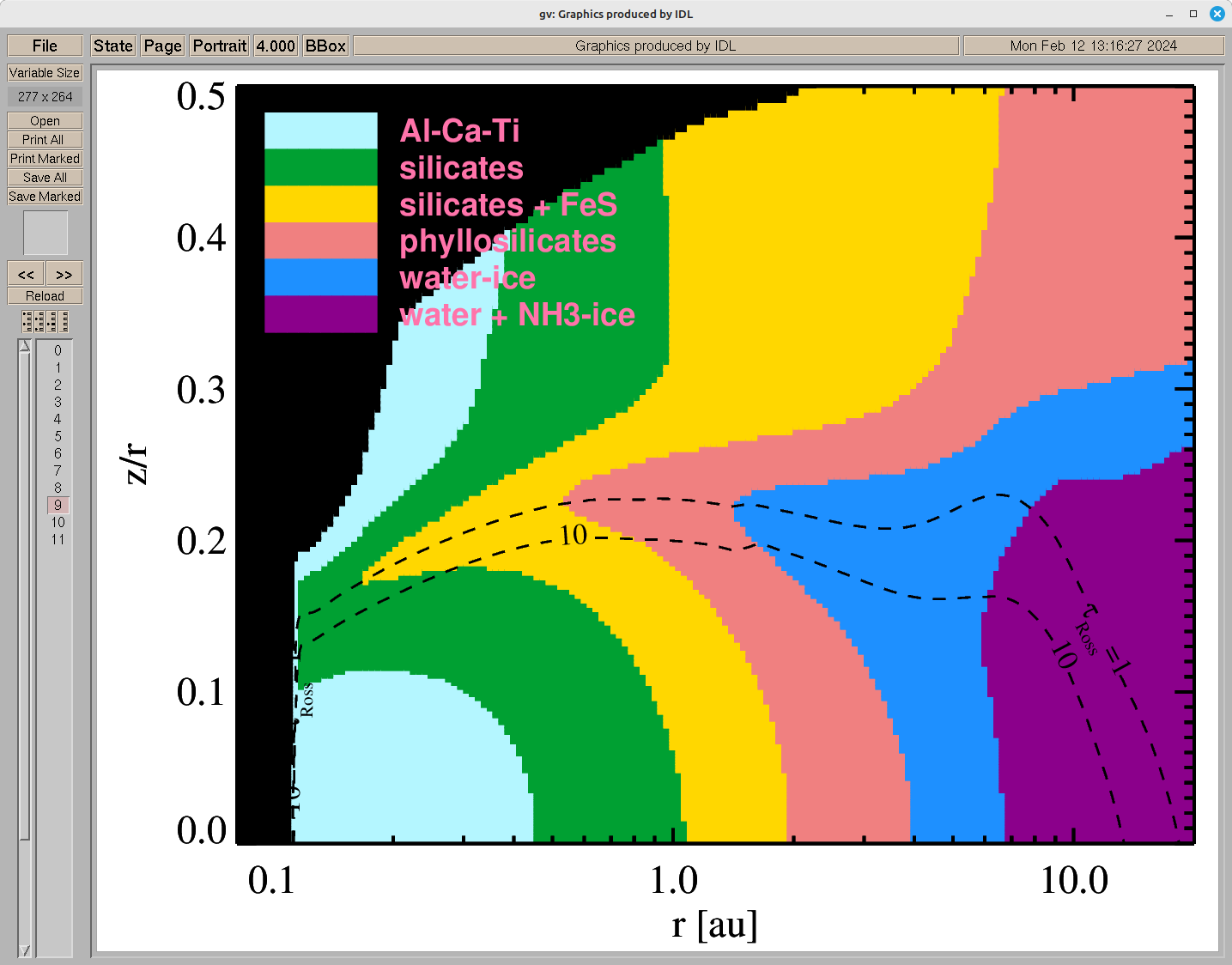} &
    \includegraphics[width=76mm,trim=143 20 20 85,clip]
                    {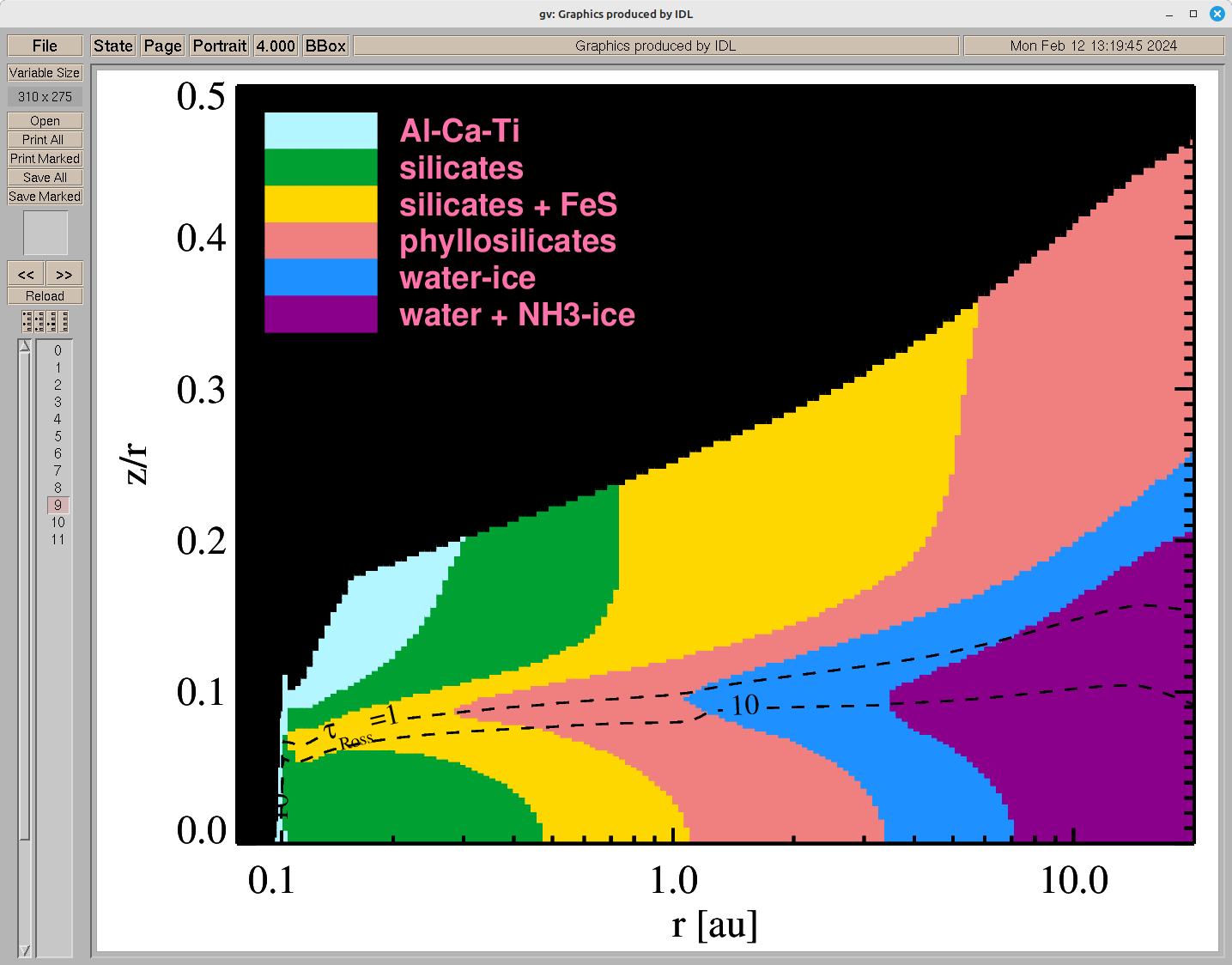} \\[-1mm]
  \end{tabular}
  \caption{Selected 2D results of disc models after stage~3 as function of radius $r$ and relative height over the midplane $z/r$.  {\bf Upper row:} hydrogen nuclei density $\nH(r,z)$.  {\bf Second row:} dust temperature structure $\Td(r,z)$.  {\bf Third row:} local dust-to-gas ratio.  {\bf Bottom row:} classes of solid mixtures$(r,z)$, see Table~\ref{tab:Vs}. Additional dashed contour lines mark the vertical Rosseland optical depth $\tau_{\rm Ross}\!=\!1$ and $\tau_{\rm Ross}\!=\!100$. The black areas in plots in the two lower rows marks the end of the GGchem modelling domain $\nH\!>\!10^7\rm\,cm^{-3}$.}
  \label{fig:big}
\end{figure*}

We start the presentation of our modelling results with Fig.~\ref{fig:Temp}, which shows some selected disc properties along the midplane, before we turn to the 2D vertical structure in Fig.~\ref{fig:big}.  Figure~\ref{fig:Temp} shows the gas and dust column densities in our models as evolutionary sequence. The earlier snapshots are similar to the first one depicted after 0.03\,Myrs.
The disc mass increases from about 0.005\,$M_\odot$ after 1000\,yrs to about 0.26\,$M_\odot$ after 1\,Mys, before it starts to decrease again, see Table~\ref{tab:star}. The radial disc extension, as measured by the radius that produces 90\% of the continuum flux at 1.3\,mm, according to the ProDiMo dust radiative transfer model, increases from about 5\,au to 130\,au.  

The dust column densities are equal to $0.004\times \Sigma(r)$ by assumption in the ProDiMo models, but Fig.~\ref{fig:Temp} shows them after dust sublimation and ice formation. In the early phases, there is an inner region ($\la\!0.5\,$au) where the silicates have sublimated in the midplane regions (only there), leading to a significant decrease of the dust column densities. One can also see a smaller, sudden increase of the dust-to-gas ratio, which is due to water ice formation.  The snowline is situated at about 3\,au after 0.03\,Myrs, 4\,au after 0.1\,Myrs, 6\,au after 0.3\,Myrs, 3\,au after 1\,Myrs, and about 1.5\,au after 3\,Myrs.

The middle column of plots in Fig.~\ref{fig:Temp} shows the calculated midplane temperatures. While the ProDiMo models (dashed cyan lines), which assume fully condensed refractory dust, result in maximum midplane temperatures as high as 4000\,K, the GGchem models (after dust sublimation) shows a temperature limit of about 1500-1700\,K that is never exceeded, indicated by the black thin dashed line: the thermostat mechanism. Fitting the GGchem results in the inner regions, where the silicates have sublimated and this limiting temperature is reached, we find 
\begin{equation}
  \ln\big(T_{\rm sil}{\rm [K]}\big) ~\approx~ \ln\big(1700\big) 
  ~+~ 0.12\,\ln\left(\frac{\Sigma\,\big[10^4{\rm\,g/cm^2}\big]}
                          {H_p\,\big[0.01{\rm\,au}\big]}\right) \ ,
  \label{eq:Tsubli}
\end{equation}
where $H_p=\big(r^3 c_T^2/GM_\star\big)^{1/2}$ is the scale height derived from the midplane sound speed $c_T$. In the following, we will denote $T_{\rm sil}$ as the silicate stability temperature.
The physical idea behind the fit in Eq.\,(\ref{eq:Tsubli}) is that the sublimation temperature increases with pressure, and $\Sigma/H_p$ is roughly proportional to the midplane gas pressure. 
The silicate stability temperature is reached inside of about 0.4\,au after 0.03\,Myrs, 0.5\,au after 0.1\,Myrs, and 0.5\,au after 0.3\,Myrs. At later evolutionary stages, the viscous heating of the disc subsides, the disc midplane becomes much cooler, and the silicate stability temperature is not reached anymore.  Beyond the snowline, the GGchem midplane temperatures (after ice formation) are slightly higher than in the ProDiMo models, because the ProDiMo models do not include ice opacities.  Consequently, the vertical optical depths are larger beyond the snowline in the GGchem models, and so the viscous heating has a stronger impact on the midplane temperature. 


The right column in Fig.\!\ref{fig:Temp} shows the computed material composition of the solid particles in terms of volume fractions in the disc midplane. The materials are only plotted when they exceed 8\% of the total solid volume anywhere in the midplane. We see a repetitive pattern of Ca-Al-Ti compounds in the innermost hot regions, followed by silicates with a complex composition.  The most important Ca-Al-Ti oxides are {\sl corundum} (\ce{Al2O3}), {\sl gehlenite} (\ce{Ca2Al2SiO7}), {\sl spinel} (\ce{MgAl2O4}), and {\sl perovskite} (\ce{CaTiO3}). There is also {\sl baddeleyite} (\ce{ZrO2}).  The most abundant silicate species are {\sl fosterite} (\ce{Mg2SiO4}), {\sl enstatite} (\ce{MgSiO3}), the feldspar end-members {\sl anthorite} (\ce{CaAl2Si2O8}) and {\sl albite} (\ce{NaAlSi3O8}), mixed with either some {\sl solid iron} at higher temperatures, or {\sl fayalite} (\ce{Fe2SiO4}) at lower temperatures.  As the midplane temperature decreases to 650\,K, sulphur starts to condense as well, mostly in form of {\sl troilite} (\ce{FeS}), before a number of phyllosilicates replace the silicates below about 350\,K, step by step, most prominently {\sl lizardite} (\ce{Mg3Si2O9H4}).  Below about 150\,K {\sl water ice} freezes out, and below about 85\,K {\sl ammonia ice} freezes out in addition.  

Figure~\ref{fig:big} shows some 2D-results of our disc model after modelling stage~3 for an early and a late evolutionary phase.  After 0.1\,Myrs, the strong viscous heating leads to a highly puffed-up inner region which puts the outer disc into a shadow. At later evolutionary phases (1\,Myrs) the inner disc flattens and we have a more textbook-like flaring disc.  The $\Td(r,z)$ structure can be subdivided into three distinct vertical layers: (1) the optically thin zone which is directly illuminated by the star, where we find vertically constant temperatures $\Td(r,z)\!\propto\!\!\sqrt{r}$, (2) a transition zone that is radially optically thick but vertically optically thin, where the temperature drops quickly as we enter the disc shadow in the downward direction, and (3) the optically thick midplane region where viscous heating leads to a temperature re-increase.    

Figure~\ref{fig:big} also shows the local dust-to-gas ratio after modelling stage~3, which does not include dust settling.  The generic value of 0.01 is only reached in the cold icy regions. The input value 0.004 is present in most other disc regions.  However, the dust-to-gas ratio drops to much lower values of order $10^{-3.5}$ to $10^{-4.5}$ in the core of the inner disc, where the silicates have sublimated; the viscous heating burns a ``hole'' into the midplane from the left, very similar to Fig.~1 in \cite{Min2011}.  However, Min et al.\ only considered one dust species.  In our models, the dust never sublimates completely, and the Al-Ca-Ti oxides remain. 

Another interesting result of our models is that the stability of the silicates is not guaranteed in the inner disc high above the midplane.  Since solid stability depends not only on temperature, but also on pressure, and the pressure drops by orders of magnitude high above the midplane, first the silicates and then even the Al-Ca-Ti oxides sublimate, and the inner disc rim takes on a rounded shape. Again, these results are very similar to Fig.~1 in \cite{Min2011}, but more smooth as we include more than just one solid species.  We have marked the vertical Rosseland optical depth $\tau_{\rm Ross}$ with contour lines for values 1 and 10 in our plots, where the latter marks the height used as boundary condition for the stage~3 modelling procedure with GGchem as explained in Sect.~\ref{sec:model3}.

The bottom row of plots in Fig.~\ref{fig:big} give an impression of the spatial distribution of the different grain materials in the disc.  We show six simplified categories of grain materials ordered by temperature: (1) the Al-Ca-Ti oxides, (2) ordinary silicates, (3) sulphur-containing silicates, (4) phyllosilicates, (5) water ice, and (6) ammonia ice.  Each category includes materials from the previous category. Examples of the complex material composition from different points of a selected disc model are listed in Table~\ref{tab:Vs}.

\begin{figure}
  \hspace*{-2mm}
  \includegraphics[page=11,width=91mm,trim=40 20 55 420,clip]{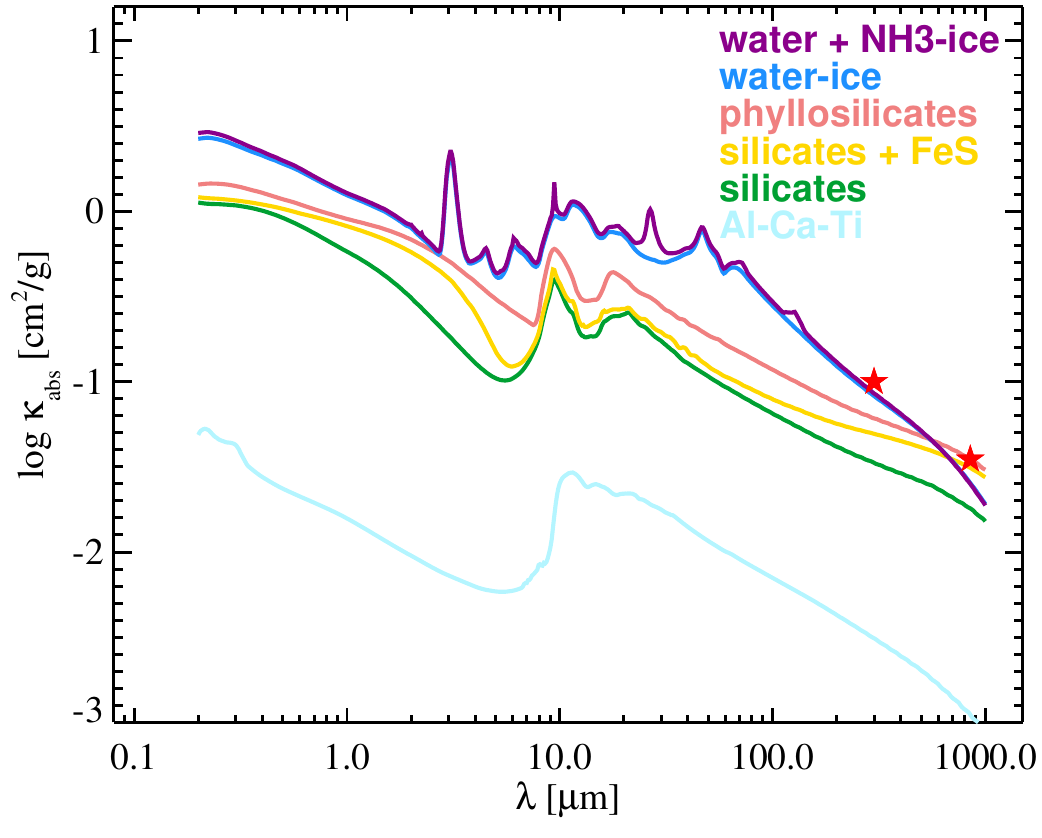}\\[-6mm]
  \caption{Calculated dust absorption opacities per gas mass for the solid mixtures typically found in our stage-3 GGchem models, compare Fig.\,\ref{fig:big}. The two red stars represent dust absorption opacities commonly used to derive disc masses from millimetre fluxes: $\rm 10\,cm^2/g(dust)$ at 1000\,GHz \citep{Beckwith1990} scaled to $\rm 3.5\,cm^2/g(dust)$ at 850\,$\mu$m.}
  \label{fig:dustopac}
\end{figure}

The absorption opacities of these categories of condensates are shown in Fig.~\ref{fig:dustopac}. We see that (1) the ordinary silicates have about one order of magnitude larger opacities than the Al-Ca-Ti oxides, (2) the inclusion of the conductive FeS into the grains leads to a substantial increase of the absorption at near-IR and millimetre wavelengths, and (3) ice formation leads to another substantial increase of the dust absorption opacities per gas mass, in particular in the UV and the far-IR, with the occurrence of additional near-IR, mid-IR and far-IR ice features.

\begin{table*}
  \caption{Material composition of the solid mixtures found in the midplane of our disc model after 0.1\,Myrs, depending on radius. The radial positions of these materials shift during disc evolution, but their relative volume compositions as function of temperature remain rather similar, compare Figs.\,\ref{fig:Temp} and \ref{fig:big}. Materials with volume fractions $<\!0.1\%$ are omitted, except for \ce{ZrO2} in the first row.}
  \label{tab:Vs}
  \vspace*{-2mm}
  \begin{tabular}{c|ccc|p{11cm}}
    \hline
    &\\[-2.3ex]
    name &  $T$\,[K]  &  \!\!\!\!$r\,([\rm au])$\!\!\!\!
         &  $\kRoss\rm\,[cm^2/g]$  &  volume composition\,[\%]\\
    &\\[-2.3ex]
    \hline
    &\\[-2ex]
    Al-Ca-Ti     & 1570  &  0.23  &  0.057  & 
                          \ce{Ca2Al2SiO7}\;[62.0],
                               \ce{Al2O3}\;[33.8],
                              \ce{CaTiO3}\;[4.2],
                                \ce{ZrO2}\;[0.01] \\[1mm]
    silicates    & 1070  &  0.64  &  0.70  &
                              \ce{MgSiO3}\;[37.2],
                             \ce{Mg2SiO4}\;[32.5], 
                                  \ce{Fe}\;[15.4],
                          \ce{CaAl2Si2O8}\;[9.7],
                           \ce{CaMgSi2O6}\;[3.1],
                             \ce{MgCr2O4}\;[1.0],
                                  \ce{Ni}\;[0.7],
                            \ce{CaTiSiO5}\;[0.3] \\[1mm]
    silicates $+$ FeS &  490  &  1.4  &  0.59  &
                             \ce{Mg2SiO4}\;[38.4],
                              \ce{MgSiO3}\;[16.7],
                                 \ce{FeS}\;[14.5],
                           \ce{NaAlSi3O8}\;[10.6],
                                  \ce{Fe}\;[8.0],
                           \ce{CaMgSi2O6}\;[6.7],
                             \ce{MgAl2O4}\;[1.0],
                             \ce{MgCr2O4}\;[0.9],
                                  \ce{Ni}\;[0.7],
                            \ce{KAlSi3O8}\;[0.7],
                        \ce{Mn3Al2Si3O12}\;[0.6],
                           \ce{Ca5P3O13H}\;[0.5],
                           \ce{Ca5P3O12F}\;[0.4],
                            \ce{CaTiSiO5}\;[0.3]\\[1mm]
    phyllosilicates & 195 &  3.1 & 0.55 &
                          \ce{Mg3Si2O9H4}\;[58.2],
                               \ce{Fe3O4}\;[10.9],
                                 \ce{FeS}\;[9.9],
                     \ce{NaMg3AlSi3O12H2}\;[8.6],
                          \ce{Fe3Si2O9H4}\;[5.3],
                        \ce{Ca3Al2Si3O12}\;[3.1],
                               \ce{Ni3S2}\;[1.0],
                      \ce{KFe3AlSi3O12H2}\;[0.7],
                               \ce{Cr2O3}\;[0.5],
                        \ce{Mn3Al2Si3O12}\;[0.5],
                                \ce{NaCl}\;[0.4],
                           \ce{Ca5P3O13H}\;[0.4],
                           \ce{Ca5P3O12F}\;[0.3],
                            \ce{CaTiSiO5}\;[0.2]\\[1mm]
    water-ice    &  119  &  4.9  &  1.00  &
                                 \ce{H2O}\;[73.7],
                          \ce{Mg3Si2O9H4}\;[15.2],
                               \ce{Fe3O4}\;[2.6],
                                 \ce{FeS}\;[2.6],
                       \ce{Mg3AlSi3O12H2}\;[2.2],
                          \ce{Fe3Si2O9H4}\;[1.4],
                        \ce{Ca3Fe2Si3O12}\;[0.9],
                               \ce{Ni3S2}\;[0.3],
                               \ce{AlO2H}\;[0.2],
                      \ce{KFe3AlSi3O12H2}\;[0.2],
                         \ce{MnAl2SiO7H2}\;[0.2],
                               \ce{Cr2O3}\;[0.1],
                                \ce{NaCl}\;[0.1]\\[1mm]
    water $+$ NH$_3$&  60  &  7.9  &  0.92  &
                                 \ce{H2O}\;[63.4],
                          \ce{Mg3Si2O9H4}\;[14.7],
                                 \ce{NH3}\;[13.5],
                               \ce{Fe3O4}\;[2.2],
                                 \ce{FeS}\;[2.2],
                          \ce{Fe3Si2O9H4}\;[1.7],
                        \ce{Ca3Al2Si3O12}\;[0.7],
                             \ce{Na2SiO3}\;[0.4],
                               \ce{AlO2H}\;[0.3],
                               \ce{Ni3S2}\;[0.3],
                      \ce{KFe3AlSi3O12H2}\;[0.2],
                               \ce{Cr2O3}\;[0.1]\\[1mm]
    \hline
  \end{tabular}
\end{table*}

\begin{figure}
  \hspace*{-2mm}
  \includegraphics[width=90mm,trim=0 0 0 0,clip]
                  {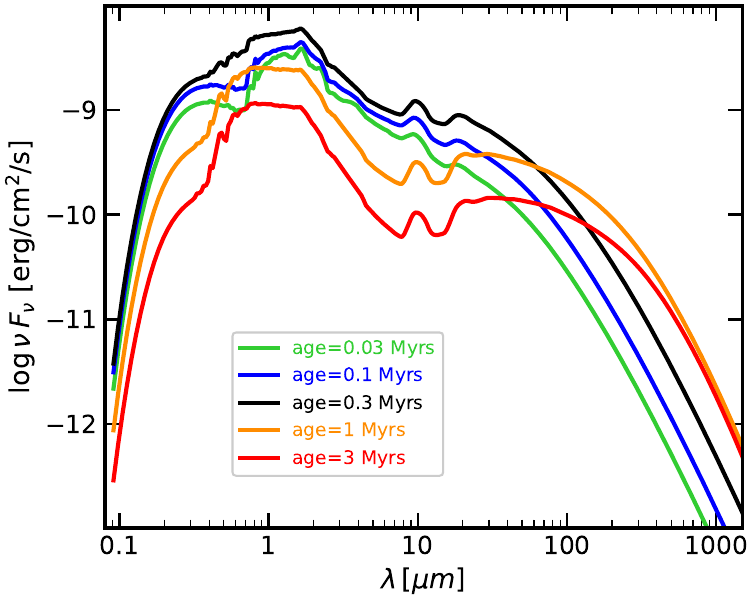}\\[-4mm]
  \caption{Spectral Energy Distributions (SEDs) as function
    of age calculated by ProDiMo (after modelling stage~2).}
  \label{fig:SEDs}
\end{figure}

Figure~\ref{fig:SEDs} shows the evolution of the spectral energy distribution (SED) over time of our disc model up to modelling stage~2. The maximum luminosity ($L_\star\!=\!3.7\,L_\odot$, $L_{\rm acc}\!=\!0.89\,L_\odot$) is reached after about 300\,000 yrs (black line), when the disc is featured by a tall "puffed-up" inner disc.  The strong internal viscous heating in this evolutionary phase causes a strong temperature inversion in the inner midplane, leading to high vertical extensions of the disc inside of about 1\,au.  This puffed-up inner disc (not an puffed-up inner rim) intersects a lot of starlight, which is then converted into near-IR and mid-IR dust emission.  In contrast, the outer disc is cold, being situated in the shadow of the tall inner disc, which leads to relatively little far-IR emission and an overall steep slope of the SED.   Once the mass accretion rate diminishes, and the viscous heating of the inner disc subsides, the inner midplane regions get a lot cooler, and the scale-heights decrease by a factor of about two.  The increase of the stellar mass with time also contributes to this evolution of the scale height, yet with less significance. 
Therefore, in the later evolutionary stages, we find a flaring disc, and a slightly warmer outer disc.  The flat inner disc intersects only little starlight, leading to much less near-IR emission up to 10\,$\mu$m.  The shape of the SED in later phases resembles transitional discs with a re-increase of the flux beyond $20\,\mu$m.

\section{Discussion}

After having established how the disc temperature structure evolves with time, and which minerals are thermodynamically stable as function of radius and time, we will now perform a timescale analysis to discuss how the grains are expected to move, grow, change internal structure, and material composition. The objective is to find a natural pathway to explain the occurrence of millimetre-sized pure Al-Ca rich grains embedded in a matrix composed of several amorphous silicate-rich materials.

\subsection{Annealing timescales}
\label{sec:tau_anneal}

Annealing is the process of internal rearrangement of a lattice structure by solid diffusion, where atoms (or groups of atoms) hop between neighbouring lattice places, which allows to purify the material and eliminate imperfections in the lattice.  The timescale for annealing according to \cite{Gail1999} is given by
\begin{equation}
  \tau_{\rm anneal}^{-1}(\,\ell\,) ~=~ 
    \frac{\lambda^2\,\nu\,\exp\Big(\frac{-E_b}{kT}\Big)}
         {3\,\ell^{\,2}} \ ,
  \label{eq:tanneal}       
\end{equation}
where $\ell$ is the total distance travelled by the atoms via a random walk, $\nu$ is a typical oscillation frequency of the lattice, and $\lambda$ is the spatial step-size, i.e.\ the distance between two neighbouring lattice places. $E_b$ is the energy barrier involved in moving the group of atoms from on lattice place to the next. For a solid layer of thickness $\ell$ to develop a well-ordered crystalline structure, as observed for the CAIs, the annealing timescale must be short compared to the residence time $t$, i.e.\ $\tau_{\rm anneal}\!\ll\!t$.  Since we are interested here in the question whether or not a particle of size $a$ can become crystalline throughout, we use $a=\ell$.

For silicates, \cite{Gail1999} estimated $\nu\!\approx\!2\times10^{13}\,$Hz from an average vibrational frequency of \ce{SiO4} tetrahedrons, and assumed $E_b/k\approx 41000\,$K based on annealing experiment of \cite{Nuth1982}, later confirmed by experiments on silicate annealing by \cite{Hallenbeck1998}.  

We consider \ce{Al2O3} as a representative material for the Al-Ca-Ti oxides.  We estimate $\nu\!\approx\!2.4\times10^{13}\,$Hz and $E_b/k\approx 48000\,$K from the general observation that annealing temperatures roughly scale with sublimation temperatures. In a solar composition gas at 1\,bar, the sublimation temperatures of \ce{Al2O3} and \ce{Mg2SiO4} have been found to be 1960\,K and 1660\,K, respectively \citep{Woitke2018}, suggesting a correction factor of 1.18. \ce{Al2O3} is known to have a hexagonal rhomboidal lattice structure with lattice constants 4.76\AA\ and 12.99\AA, so the annealing becomes dependent on direction, also because $E_b$ is likely to depend on direction.  However, here we simply assume $\lambda\!\approx\!1\,$nm. 
The resulting annealing timescales are shown in Fig.~\ref{fig:timescales} for different particle sizes $a$, and will be discussed together with the coagulation timescales at the end of the next section. 



\subsection{Coagulation timescales}
\label{sec:tau_coag}

A particle of size $a$ will undergo $\pi\,(a+a')^2\Delta v\;\nH\,f(a')\,da'$ collisions per second with particles of sizes $a'...\,a'\!+da'$, where $f(a)$ is the dust size distribution function [$\rm cm^{-1}/$H-nucleus]. $\Delta v(a,a')$ is the turbulence-induced relative velocity between particles of sizes $a$ and $a'$, see \cite{Ormel2007}, and $\pi(a+a')^2$ is the geometrical cross section.  We define the timescale of coagulation as the time required to increase a particle's volume (or mass) by a factor of $\exp(1)$  
\begin{equation}
  \tau_{\rm coag}^{-1}(a) ~=~ \frac{\frac{dV}{dt}}{V}~=~ \nH \int\!\min\Big\{1,\frac{a'^3}{a^3}\Big\}\;\pi\,(a\!+\!a')^2\,\Delta v\;\alpha_S f(a')\;da' \,,
  \label{eq:tcoag}
\end{equation}
where $V\!=\!(4\pi/3)\,a^3$ is the particle volume and $(4\pi/3)\,a'^3$ is the increase of the particle's volume by one sticking collision. $\alpha_S$ is the sticking efficiency. We use $\min\!\big\{1,a'^3/a^3\big\}$ in the integration kernel instead of just $a'^3/a^3$ to avoid interpreting the case $dV>V$ as super fast growth.  Instead, in the case of small grains colliding with big grains ($a\!<\!a'$), we use just their collisional frequency.  

To calculate the turbulence-induced relative velocities $\Delta v(a,a')$, we use the semi-empirical formulae of \cite{Ormel2007}, which depend on the local sound speed, the turbulence parameter $\alpha$, the two Stokes numbers of the colliding particles, and the gas Reynolds number. At gas densities of about $10^{15}\rm\,cm^{-3}$, typical for these midplane regions, the Stokes numbers of mm-sized particles are of order $\St\!\approx\!10^{-4}$, so we mostly have the limiting case of small Stokes numbers (small particles) as described by equation (26) in \cite{Ormel2007}, where $\Delta v\approx c_T\,(3\,\alpha\,\St)^{1/2}$, see also \cite{Birnstiel2023}. 
In that case, close inspection of Eq.\,(\ref{eq:tcoag}) shows that for all sizes $a$, the integral on the right side is dominated by the largest collisional partners as these collisions have the largest $\Delta v$. 
For small particles $a$, the coagulation timescale turns out to be size-independent, whereas for the largest particles, it is slightly longer due to the lack of even larger particles.  

In the close midplane regions, where the silicates sublimate, we find a dust-to-gas mass ratio of approximately $2\times10^{-4}$, corresponding to a size-reduction factor of $\gamma\!\approx\!0.3$ (see Eq.\,\ref{eq:gamma}).  Therefore, according to our assumption $\amax\!=\!3$mm, the largest Al-Ca-Ti oxide particles are initially about 1\,mm in size after silicate sublimation, and with $\alpha\!=\!10^{-3}$, we find relative velocities $\Delta v$ between any particles and mm-sized particles between a few 10\,cm/s to a few m/s.

These relative velocities are already close to the fragmentation threshold velocity \citep{Blum2018,Birnstiel2023}. Furthermore, for large particles colliding with large particles, the sticking efficiency is likely to be much smaller than $\alpha_S\!=\!1$ because of the bouncing barrier \citep[e.g.,][]{Guttler2010}. When mm-sized particles collide with mm-sized or even larger particles, the contact (van der Waal's) forces during the collisions become negligible in comparison to their inertia, and therefore, such particles simply do not stick together when they collide, they just bounce. For simplicity, we calculate the integral in Eq.\,(\ref{eq:tcoag}) anyway up to $\amax$ (few millimetres), with $\alpha_S\!=\!1$, to get an order of magnitude estimation of the coagulation timescale, keeping in mind that the actual timescales can be significantly longer because of the fragmentation threshold and the bouncing barrier.

\begin{figure}
  \hspace*{-1mm}
  \includegraphics[width=90mm,trim=15 15 10 15,clip]
                  {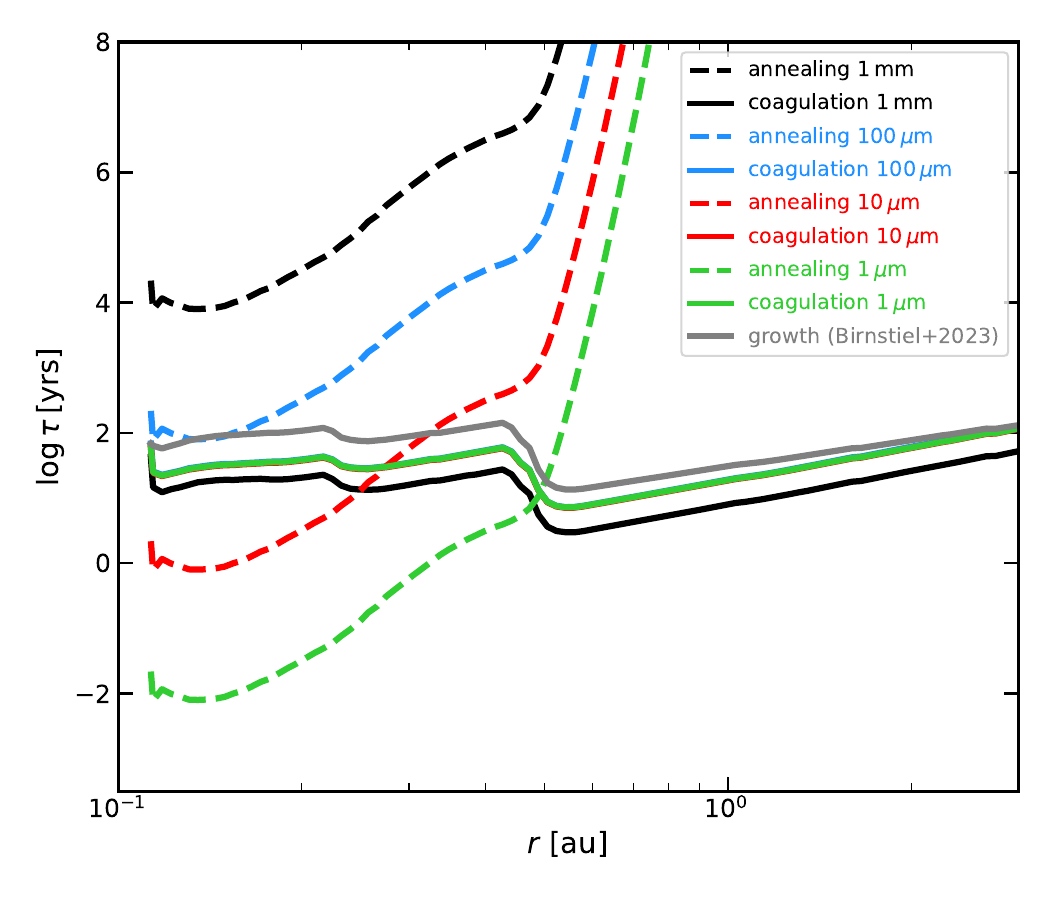}\\*[-5mm]
  \caption{Annealing and coagulation timescales for Al-Ca-Ti oxide particles of different sizes in the disc midplane as function of $r$ for an age of the solar nebula of 0.1\,Myrs. The silicates have sublimated inside of about 0.45\,au. Inside of about 0.2\,au, only \ce{Al2O3} and \ce{ZrO2} remain, whereas outside of 0.2\,au, \ce{Ca2Al2SiO7} and \ce{CaTiO3} are present as well, which creates another small step in the dust-to-gas ratio and hence in the coagulation timescales.}
  \label{fig:timescales}
\end{figure}

Our timescale analysis is shown in Fig.~\ref{fig:timescales} for the disc model after 0.1\,Myrs. We compare our coagulation timescale (Eq.~\ref{eq:tcoag}) with the simplified growth timescale proposed by \cite{Birnstiel2023}
\begin{equation}
  \tau_{\rm growth}^{-1} = Z\,\Omega \ ,
  \label{eq:taugrowth}
\end{equation}
where $Z$ is the dust-to-gas mass ratio and $\Omega$ is the Keplerian rotation frequency, showing good agreement. 
Thus, creating mm-sized particles by coagulation is a fast process taking only about 100\,yrs. The growth by coagulation will stop when the particles have reached the fragmentation barrier. 
Once the particles have reached this barrier, fragmentation replenishes the small particles \citep{Blum2008,Birnstiel2023}, and an equilibrium between coagulation and fragmentation establishes a power-law size distribution up to the fragmentation barrier.  

In the close midplane, where temperatures are regulated to remain close to the silicate sublimation temperatures $\approx\!1500-1700\,$K (Eq.~\ref{eq:Tsubli}), annealing a 1\,mm Al-Ca-Ti oxide particle takes about $10^4\,$yrs, which is still short compared to the lifetime $\sim\!10^5\,$yrs of the $T$-regulated midplane zone, where the silicates are predominantly in the gas phase.  However, already beyond 0.3\,au, where the temperatures are just slightly lower, $1400-1500\,$K, the annealing timescales increase significantly.  However, it is still possible to form 100\,$\mu$m annealed Al-Ca-Ti oxide particles there, which can then coagulate quickly. 

The situation changes substantially outside of about 0.45\,au in this model after 0.1\,Myrs, where the silicates are thermally stable and the thermostat mechanism breaks down.  The temperature drops quickly, the dust-to-gas ratio increases by almost 2 orders of magnitude, the coagulation timescales (Eq.\,\ref{eq:taugrowth}) become even shorter, and the annealing timescales become huge.  In this case, we expect the silicates to re-condense on the existing Al-Ca-Ti oxide surfaces to form an amorphous mantle mainly composed of \ce{Mg2SiO4} and \ce{MgSiO3}, without having the time to form an ordered crystal structure.  Other materials like feldspar-components, iron and nickel, and later FeS, will be incorporated in this mantel, see Table~\ref{tab:Vs}.  These core-mantle grains will continue to collide, fragment and coagulate.  The expected physical properties of this mantel correspond well to the ``matrix'' as observed in chondritic materials.
Laboratory analysis has shown that the matrix contains olivines, pyroxenes, Fe-Ni metals, lesser amounts of sulphides, sulphates, carbonates, and in some cases significant amounts of phyllosilicates \citep{Brearley1989, Busek1993, Scott2005}.

\subsection{Range of validity of the thermostat mechanism}
\label{sec:regulate}

As discussed in Sect.\ref{sec:opacities}, the thermostat regulation mechanism enforces the midplane temperature to stay close to the silicate sublimation temperature (Eq.\,\ref{eq:Tsubli}), provided that the viscous heating is strong enough and the disc is optically thick enough.  However, there is a limit of this regulation mechanism, namely when the dust opacity becomes smaller than the gas opacity.  In that case, a further increase of the local temperature does not result in a significant lowering of the total Rosseland opacity (Eq.\,\ref{eq:kRossTot}), and hence the thermostat breaks down.

\begin{figure}
  \vspace*{1mm}
  \hspace*{-1mm}
  \includegraphics[page=1,width=90mm,trim=15 57 10 15,clip]
                  {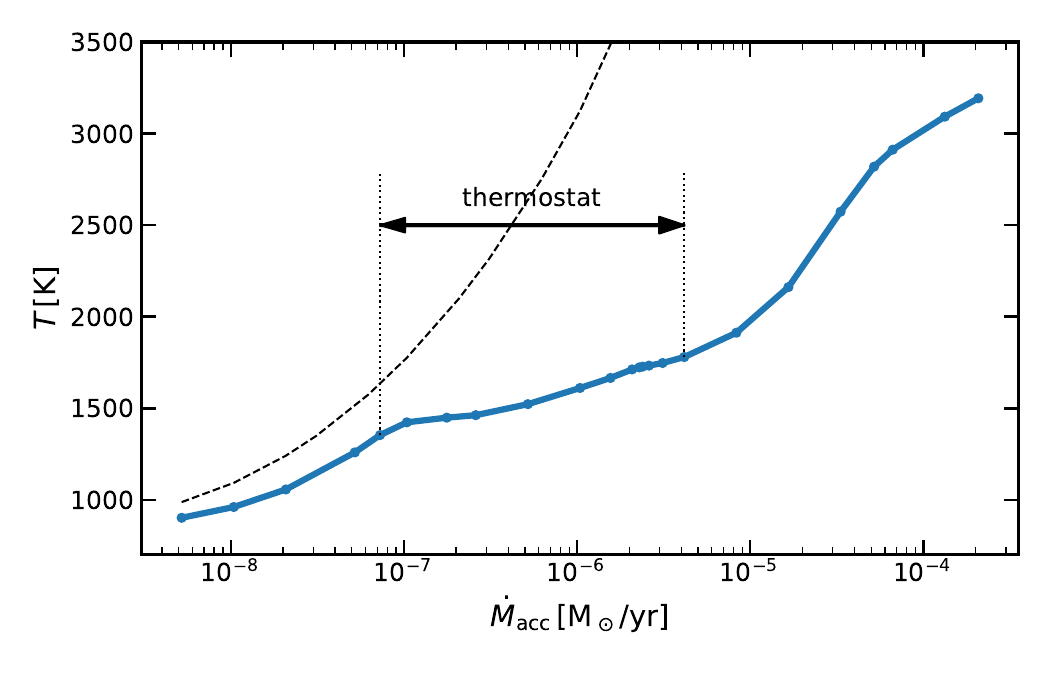}\\*[-0.5mm]
  \includegraphics[page=2,width=90mm,trim= 5.5 15 9 15,clip]
                  {Figs/FvisPlot.pdf}\\*[-5mm]
  \caption{Range of validity of the thermostat regulation mechanism. The upper plot shows the calculated midplane temperatures in our disc model at $r\!=\!0.2\,$au after 0.1\,Myrs for a sequence of simulations where we modified the mass accretion rate $\dot{M}_{\rm acc}$ from its nominal value of $1.04\times10^{-6}\rm\,M_\odot/yr$. The dashed line shows the resulting midplane temperatures when we use the fixed dust opacities from the ProDiMo model. The lower plot shows the corresponding dust and gas Rosseland opacities in the disc midplane, taking into account dust sublimation.}
  \label{fig:RangeOfValidity}
\end{figure}

In order to study the range of validity of the thermostat mechanism, we calculated a series of stage-3 (GGchem) models where we artificially modified the viscous heating rate $F_{\rm vis}$ (Eq.~\ref{eq:Fvis}) by a factor ranging from 0.005 to 200, see Fig.~\ref{fig:RangeOfValidity}. We selected a representative column at $r\!=\!0.2\,$au from our disc model after 0.1\,Myrs for this investigation.  Since $F_{\rm vis}$ is proportional to the mass accretion rate, we plot the results over $\dot{M}_{\rm acc}$. The nominal values of the unmodified model are $\dot{M}_{\rm acc}\!=\!1.04\times10^{-6}\rm\,M_\odot/yr$ and $F_{\rm vis}\!=\!4.14\times10^6\rm\,erg/cm^2$, which corresponds to the centre of the plot, where we reach $T\!\approx\!1600\,$K in the midplane.

\begin{figure*}[!t]
  \hspace{-2.5mm}
  \begin{tabular}{cc}
    {\sf Scenario 1} & {\sf Scenario 2} \\[0mm]
    \includegraphics[width=90mm,trim=0 0 0 0,clip]{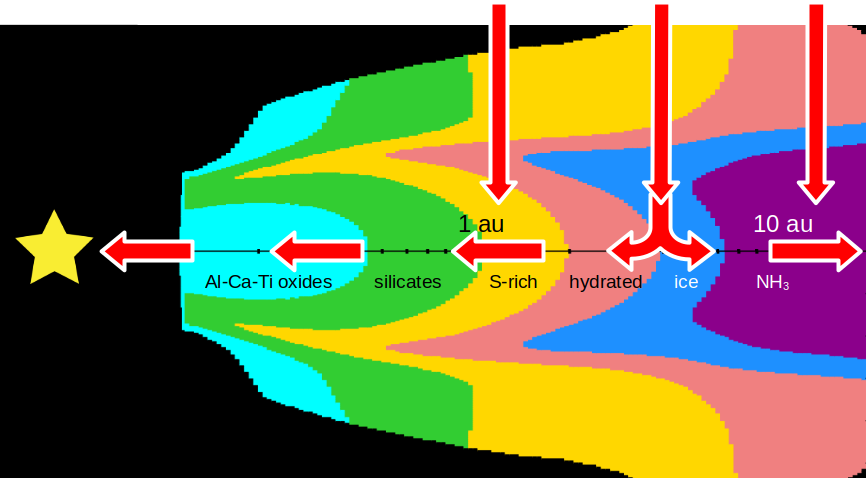} &
    \hspace*{-2mm}
    \includegraphics[width=90mm,trim=0 0 0 0,clip]{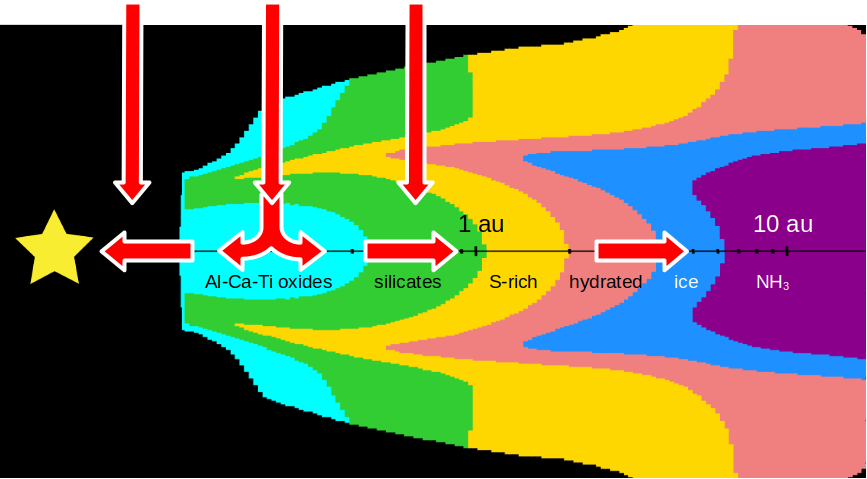} \\[-1mm]
  \end{tabular}
  \caption{Two scenarios for the early evolution of the solar nebula. In the common scenario~1, the disc is fed by the molecular cloud at relatively large radial distances, and the gas and the dust grains move inwards.  In our favoured scenario~2, the disc is fed at much smaller radii, and the dust grains are dragged outward with the viscously spreading gas, which provides a natural explanation for the formation of CAIs embedded in a silicate matrix.}
  \label{fig:scenario}
\end{figure*}

As we increase $\dot{M}_{\rm acc}$ from very low values, we reach the silicate sublimation temperature at about $7\times10^{-8}\rm\,M_\odot/yr$, beyond which the dust opacity $\kappa_{\rm R}^{\rm dust}$ decreases rapidly, until only the Al-Ca-Ti oxides remain in the midplane.  This dust opacity, however, is still larger than the gas opacity by about one order of magnitude.  Increasing $\dot{M}_{\rm acc}$ further, to a critical value of about $4\times10^{-6}\rm\,M_\odot/yr$, has little effect on the midplane temperature; this is the range where the thermostat mechanism is at work.  However, for even larger $\dot{M}_{\rm acc}$, the Al-Ca-Ti oxides eventually start to sublimate, too, and once their opacity becomes negligible, the gas opacity takes over the regulation of the further increase of the midplane temperature with increasing $\dot{M}_{\rm acc}$.  

Therefore, we expect the Al-Ca-Ti oxides to sublimate when the mass accretion rate exceeds the order of $\approx\!10^{-6}$ to $10^{-5}\rm\,M_\odot/yr$.   The exact value of that critical mass accretion rate will depend on the surface density, which is about $10^{\,4}\rm\,g/cm^2$ in our model at 0.2\,au after 0.1\,Myrs.  If the surface density is smaller, the optical depths are smaller, leading to less pronounced temperature inversions, and the critical mass accretion rate will be higher.

\subsection{Shortcomings of the model, and future improvements}
\label{sec:shorts}

Although our model gives a clear picture of how dust grains move in the early solar nebula, what their thermal history is, which we have determined using detailed radiative transfer calculations, and how the different minerals sublimate and re-sublimate as a function of time, we are far from a fully consistent, time-dependent 3D treatment of the CAI formation problem. In each modelling stage (Fig.~\ref{fig:ModelStages}), we made slightly different assumptions about dust opacities and temperatures that could have an impact on the previous modelling stages.  For example, in the second modelling stage (ProDiMo) we assume a constant dust material mixture with a settled size distribution between constant minimum and maximum grain sizes, whereas in the third modelling stage (GGchem) we assume a varying material composition and an unsettled powerlaw size distribution with adjusted maximum and minimum size according to the resulting dust/gas mass ratio.  However, these differences only play a role in the optically thick hot core of the disk where the silicates can sublimate.  We checked that at the boundary $\tau_{\rm Ross}\!\sim\!10$, below which the GGchem models are applied, the Rosseland opacities in both models agree with each other to within a factor of two. An implementation of the dust opacity including dust sublimation in the phase-1 hydro-simulatiomns might be feasible in future applications, but goes beyond the scope of this paper. 

The midplane temperatures (see Fig.~\ref{fig:Temp}) can indeed change substantially when silicate sublimation is taken into account, but this is relevant only in the optically thick midplane ($\tau_{\rm Ross}\!>\!10$). The temperature deviations in those deep layers, however, no not affect the upper layers, as the upper layers only see the boundary temperature, which remains unchanged in modelling phase 3.  There is one exception though, and that is the vertical hydrostatic structure, which is more extended when the midplane is warmer.  Here, when we compute the hydrostatic vertical structure in ProDiMo, we have implemented a cap of the local dust temperature by the silicate sublimation temperature (Eq.\,\ref{eq:Tsubli}) to make sure we do not overpredict the vertical scale heights.

In addition we note that all dust stability computations with GGchem are time-independent by design.  They are based on the principle of minimisation of the total Gibbs free energy, so we assume that all internal rearrangement processes necessary to reach the thermodynamically most favourable solid state happen in due time, which becomes increasingly questionable when the temperatures become low, see \citet[][their Fig.\,11]{Herbort2020}. However, close to the stability temperatures of all minerals, sublimation and re-sublimation are actually quite fast processes, and we have the annealing timescales to discuss the rearrangements.  In fact, we do not follow the grains in a time-dependent way and calculate how their position, size, material composition and internal structure change, but we base all our conclusions on the trajectories found in the first modelling phase, and the condensation, growth, and annealing timescales found in third modelling phase, assuming that an equilibrium between dust coagulation and fragmentation holds and always installs a powerlaw size distribution quickly. A more detailed treatment of these processes in the frame of the phase-1 hydrodynamical disc evolution model would be desirable.

\section{Summary and conclusions}
This paper proposes a pathway to form millimetre-sized, pure, and annealed Al-Ca-Ti oxide grains embedded in a matrix of amorphous silicates in the earliest evolutionary phases of the solar nebula.  We associate these particles with the Cal-Al-rich inclusions (CAIs) found in many chondritic meteorites.

Our model combines (1) the 1D viscous disc evolutionary models of \citet{Drazkowska2018,Drazkowska2023b} with (2) the 2D radiation thermo-chemical disc models of \citet{Woitke2024}, (3) the equilibrium condensation models of \cite{Woitke2018}, and (4) new dust opacity calculations.  

The models reveal a thermostat mechanism in the disc midplane regions close to the star \citep{Min2011}, which keeps the local temperature stable at about $\rm 1500-1700$\,K, just above the silicate sublimation temperature.  All dust grains inherited from the molecular cloud, which temporarily reside in this region, will loose their silicates and all other more volatile components quickly, making the Al-Ca-Ti oxides the only condensates that remain in this region. The thermostat mechanism works by lowering the dust opacity by silicate sublimation, until the viscous heat accumulating in the midplane can escape vertically.

These particular, regulated temperature conditions remain relatively stable for hundreds of thousands of years, which allow the Al-Ca-Ti particles to grow by coagulation and create a semi-ordered crystal structure by annealing. The growth of these particles by coagulation only stops once the size of the Al-Ca-Ti particles reaches the fragmentation barrier of a few millimetres; creating a size distribution that we can still observe today by measuring the sizes of the CAIs in chondritic materials, see e.g.\ \cite{Nakamura2007}.

To form the CAIs, i.e.\ to embed these particles in a silicate matrix, it is essential to have an outward motion of the dust grains during the earliest evolutionary phases, as is the case in the \citet{Drazkowska2018,Drazkowska2023b} models, which is sketched as scenario~2 on the right side of Fig.~\ref{fig:scenario}.  The Al-Ca-Ti particles are dragged out with the viscously expanding gas to radii beyond 0.5\,au, where the thermostat regulation mechanism breaks down, and the silicates re-condense on the existing surfaces of the Al-Ca-Ti oxide particles.  When this happens, 
temperatures are already too low for annealing, so this created an amorphous silicate mantel.  In contrast, according to the more common scenario~1 in Fig.~\ref{fig:scenario}, where all matter is accreted outside-in through the disc, at no time and space in the disc one would expect to find large, pure and crystalline Al-Ca-Ti rich particles embedded in an amorphous silicate matrix.

\begin{figure}[!t]
  \vspace*{-2mm}
  \centering
  \includegraphics[width=87mm,trim=0 0 0 0,clip]{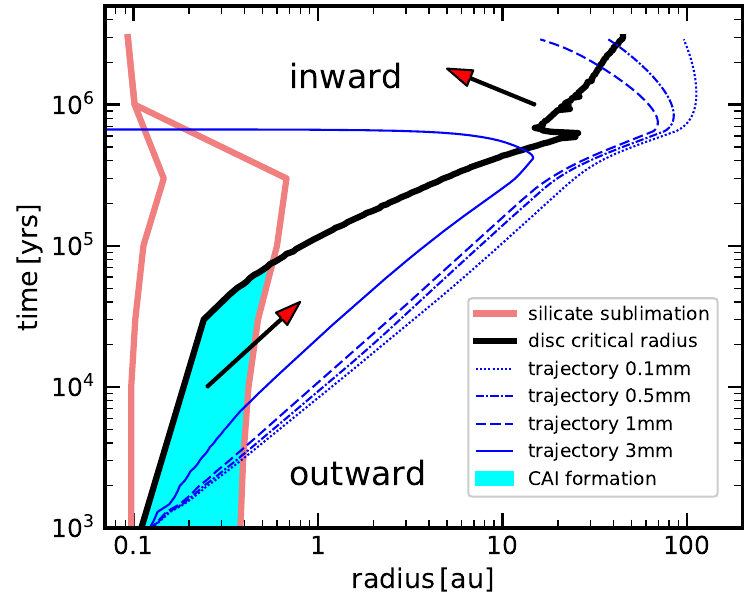}\\[-1mm]
  \caption{CAI formation in time and space. The critical radius (black line) divides this diagram into a region where millimetre grains move inward (due to inward gas motion and radial drift) and a region where these grains are dragged outward with the expanding gas. The light red line encircles the hot $T$-regulated region with thermally unstable silicates, leading to pure Al-Ca-Ti oxide particles.  The cyan shaded area is the region of lasting CAI-formation. The Al-Ca-Ti oxide particles formed in this region will eventually reside in a silicate matrix and populate the disc. Four trajectories for different particle sizes are plotted with thin blue lines.}
  \label{fig:overlap}
\end{figure}

Our scenario for the CAI-formation in the early solar system hence requires two ingredients: (1) the existence of an inner disc zone where the temperature is regulated to stay just above the silicate sublimation temperature, and (2) an outward flow of solid particles through this zone.  According to our disc model, both requirements are met, as shown in Fig.\,\ref{fig:overlap} by the cyan overlap region, up to a disc age of about 50\,000\,yrs.  
Eventually, after the mass accretion rate has fallen to values $\la\!10^{-7}\rm\,M_\odot/yr$ and the viscous heating of the disc subsides, the thermostat mechanism breaks down, 
and the dust particles from the molecular cloud that continue to join the disc will do so mostly unaltered, i.e.\ without annealing and without losing their silicates of other refractory components on the way.

\begin{acknowledgements}
   P.\,W.\ acknowledges funding from the European Union
   H2020-MSCA-ITN-2019 under Grant Agreement no.\,860470 (CHAMELEON).
   J.\,D.\ acknowledges funding from the European Union under the European Union’s Horizon Europe Research \& Innovation Programme 101040037 (PLANETOIDS). 
   P.\,M.\ acknowledges funding from the Italian Ministerial Grant PRIN 2022, “Radiative opacities for astrophysical applications”, no. 2022NEXMP8, CUP C53D23001220006.

\end{acknowledgements}

\bibliographystyle{aa}
\bibliography{references}

\appendix
\section{Optical constants}

Table~\ref{tab:opac} lists the names, chemical formula and references for optical constants for the solid materials that occur frequently and abundantly in our disc model.

\begin{table*}
\caption{Solid materials and optical constants. "am." means amorphous. "$\to$" means a replacement.}
\label{tab:opac}
\vspace*{-2mm}\hspace*{-1mm}
\resizebox{!}{121mm}{
\begin{tabular}{c|c|l}
\hline
&&\\[-2.2ex]
material & sum formula & references for optical constants / replacements\\[0.2ex]
\hline
\multicolumn{3}{l}{ }\\[-2.1ex]
\multicolumn{3}{l}{\hspace*{27mm}\bf high-temperature condensates}\\[0.2ex]
\hline
&&\\[-2.2ex]
solid tungsten      & \ce{W}              & \cite{Ordal1988} \\
baddeleyite         & \ce{ZrO2}           & \cite{Dowling1977} \\
corundum            & \ce{Al2O3}          & \cite{Begemann1997} \\
perovskite          & \ce{CaTiO3}         & \cite{Posch2003} \\
gehlenite           & \ce{Ca2Al2SiO7}     & \cite{Mutschke1998} \\
akermanite          & \ce{Ca2MgSi2O7}     & $\to$ \ce{Ca2Al2SiO7} \\
\hline
\multicolumn{3}{l}{ }\\[-2.1ex]
\multicolumn{3}{l}{\hspace*{27mm}\bf silicates}\\[0.2ex]
\hline
&&\\[-2.2ex]
am.\,fosterite      & \ce{Mg2SiO4}        & \cite{Jaeger2003} \\
am.\,enstatite      & \ce{MgSiO3}         & \cite{Jaeger2003} \\
am.\,silica         & \ce{SiO2}           & \cite{Henning1997} \\
silicon monoxide    & \ce{SiO}            & \cite{Wetzel2013} \\
jadeite             & \ce{NaAlSi2O6}      & \cite{Mutschke1998} \\
larnite             & \ce{Ca2SiO4}        & $\to$ \ce{Mg2SiO4} \\
picrochromite       & \ce{MgCr2O4}        & $\to$ \ce{MgFeSiO4} \\
spessartine         & \ce{Mn3Al2Si3O12}   & $\to$ \ce{MgFeSiO4} \\
grossular           & \ce{Ca3Al2Si3O12}   & $\to$ \ce{MgFeSiO4} \\
nepheline           & \ce{NaAlSiO4}       & $\to$ \ce{MgFeSiO4} \\
kalsilite           & \ce{KAlSiO4}        & $\to$ \ce{MgFeSiO4} \\
fayalite            & \ce{Fe2SiO4}        & \cite{Fabian2001} \\
wollastonite        & \ce{CaSiO3}         & \cite{Shaker2018} \\
sphene              & \ce{CaTiSiO5}       & $\to$ \ce{CaTiO3} \\
diopside            & \ce{CaMgSi2O6}      & \cite{Arnold2014} \\
jadeite             & \ce{NaAlSi2O6}      & \cite{Mutschke1998} \\
kosmochlor          & \ce{NaCrSi2O6}      & $\to$ \ce{CaMgSi2O6} \\
andradite           & \ce{Ca3Fe2Si3O12}   & $\to$ Mg$_{0.7}$Fe$_{0.3}$SiO$_3$ \\
zirconium silicate  & \ce{ZrSiO4}         & $\to$ \ce{ZrO2} \\
\hline
\multicolumn{3}{l}{ }\\[-2.1ex]
\multicolumn{3}{l}{\hspace*{27mm}\bf feldspar components}\\[0.2ex]
\hline
&&\\[-2.2ex]
albite              & \ce{NaAlSi3O8}      & \cite{Mutschke1998} \\
anorthite           & \ce{CaAl2Si2O8}     & $\to$ \ce{NaAlSi3O8} \\
microcline          & \ce{KAlSi3O8}       & $\to$ \ce{NaAlSi3O8} \\
\hline
\multicolumn{3}{l}{ }\\[-2.1ex]
\multicolumn{3}{l}{\hspace*{27mm}\bf metals and metal oxides}\\[0.2ex]
\hline
&&\\[-2.2ex]
solid nickel        & \ce{Ni}             & \cite{Ordal1987} \\
solid copper        & \ce{Cu}             & \cite{Ordal1985} \\
solid iron          & \ce{Fe}             & \cite{Palik1991} \\
ferropericlase      & \ce{FeO}            & \cite{Henning1995} \\
hematite            & \ce{Fe2O3}        
  & \small \url{https://www.astro.uni-jena.de/Laboratory/OCDB/mgfeoxides.html} \\
magnetite           & \ce{Fe3O4}          
  & \small \url{https://www.astro.uni-jena.de/Laboratory/OCDB/mgfeoxides.html} \\
ilmenite            & \ce{FeTiO3}         & $\to$ \ce{Fe2O3} \\
hercynite           & \ce{FeAl2O4}        & $\to$ \ce{MgAl2O4} \\
spinel              & \ce{MgAl2O4}        & \cite{Zeidler2011} \\
periclase           & \ce{MgO}            & \cite{Palik1991} \\
rutile              & \ce{TiO2}           &  \cite{Zeidler2011} \\
titanium oxide      & \ce{Ti4O7}          & $\to$ \ce{TiO2} \\
titanium oxide      & \ce{Ti3O5}          & $\to$ \ce{TiO2} \\
eskolaite           & \ce{Cr2O3}          & $\to$ \ce{Fe2O3} \\
\hline
\multicolumn{3}{l}{ }\\[-2.1ex]
\multicolumn{3}{l}{\hspace*{27mm}\bf materials containing sulphur}\\[0.2ex]
\hline
&&\\[-2.2ex]
troilite            & \ce{FeS}            & \cite{Henning1997} \\
heazlewoodite       & \ce{Ni3S2}          & $\to$ \ce{FeS} \\
sodium sulfide      & \ce{Na2S}           & \cite{Khachai2009}\\
sphalerite          & \ce{ZnS}            & \cite{Palik1991} \\
alabandite          & \ce{MnS}            & \cite{Huffman1967} \\
\hline
\multicolumn{3}{l}{ }\\[-2.1ex]
\multicolumn{3}{l}{\hspace*{27mm}\bf phyllosilicates}\\[0.2ex]
\hline
&&\\[-2.2ex]
lizardite           & \ce{Mg3Si2O9H4}     & $\to$ Mg$_{0.7}$Fe$_{0.3}$SiO$_3$ \\
sodaphlogopite      & \ce{NaMg3AlSi3O12H2}& $\to$ Mg$_{0.7}$Fe$_{0.3}$SiO$_3$ \\
greenalite          & \ce{Fe3Si2O9H4}     & $\to$ Mg$_{0.7}$Fe$_{0.3}$SiO$_3$ \\
phlogopite          & \ce{KMg3AlSi3O12H2} & $\to$ Mg$_{0.7}$Fe$_{0.3}$SiO$_3$ \\
lawsonite           & \ce{CaAl2Si2O10H4}  & $\to$ Mg$_{0.7}$Fe$_{0.3}$SiO$_3$ \\
annite              & \ce{KFe3AlSi3O12H2} & $\to$ Mg$_{0.7}$Fe$_{0.3}$SiO$_3$ \\
talc                & \ce{Mg3Si4O12H2}    & $\to$ Mg$_{0.7}$Fe$_{0.3}$SiO$_3$ \\
Mn-chloritoid       & \ce{MnAl2SiO7H2}    & $\to$ Mg$_{0.7}$Fe$_{0.3}$SiO$_3$ \\
diaspore            & \ce{AlO2H}          & $\to$ \ce{Al2O3} \\
goethite            & \ce{FeO2H}          & $\to$ \ce{FeO} \\
brucite             & \ce{MgO2H2}         & $\to$ \ce{MgO} \\
\hline
\multicolumn{3}{l}{ }\\[-2.1ex]
\multicolumn{3}{l}{\hspace*{27mm}\bf other materials}\\[0.2ex]
\hline
&&\\[-2.2ex]
am.\,carbon         & \ce{C}              & \cite{Zubko1996} \\
dolomite            & \ce{CaMgC2O6}       
    & \small \url{https://apps.dtic.mil/sti/citations/ADA192210} \\
halite              & \ce{NaCl}           & \cite{Palik1991} \\
sylvite             & \ce{KCl}            & \cite{Palik1991} \\
water ice           & \ce{H2O}            & \cite{Warren2008} \\
amonia ice          & \ce{NH3}            & \cite{Martonchik1984} \\
\hline
\multicolumn{3}{l}{ }\\[-2.1ex]
\multicolumn{3}{l}{\hspace*{27mm}\bf replacement materials}\\[0.2ex]
\hline
&&\\[-2.2ex]
am. olivine         & \ce{MgFeSiO4}       & \cite{Dorschner1995} \\
am. pyroxene        & Mg$_{0.7}$Fe$_{0.3}$SiO$_3$ & \cite{Dorschner1995} \\
\hline
\end{tabular}}\\[2mm]
{\tiny
Electronic files containing the optical constants (real and imaginary parts of the refractory index $(n,k)$) can be\\ collected from 
\url{https://zenodo.org/records/8221362} and \url{https://github.com/cdominik/optool}.}
\end{table*}

\section{Phasediagrams for Al$_2$O$_3$ and SiO$_2$}
\label{sec:phasedia}

\begin{figure*}
  \vspace*{1mm}
  \begin{tabular}{ccc}
  &{\sf hydrogen-rich} & {\sf hydrogen-poor}\\
  \begin{minipage}{6mm}{\sf SiO$_2$}:\\*[70mm]\end{minipage} &
  \includegraphics[page=3,width=80mm,trim=0 0 0 0,clip]
                  {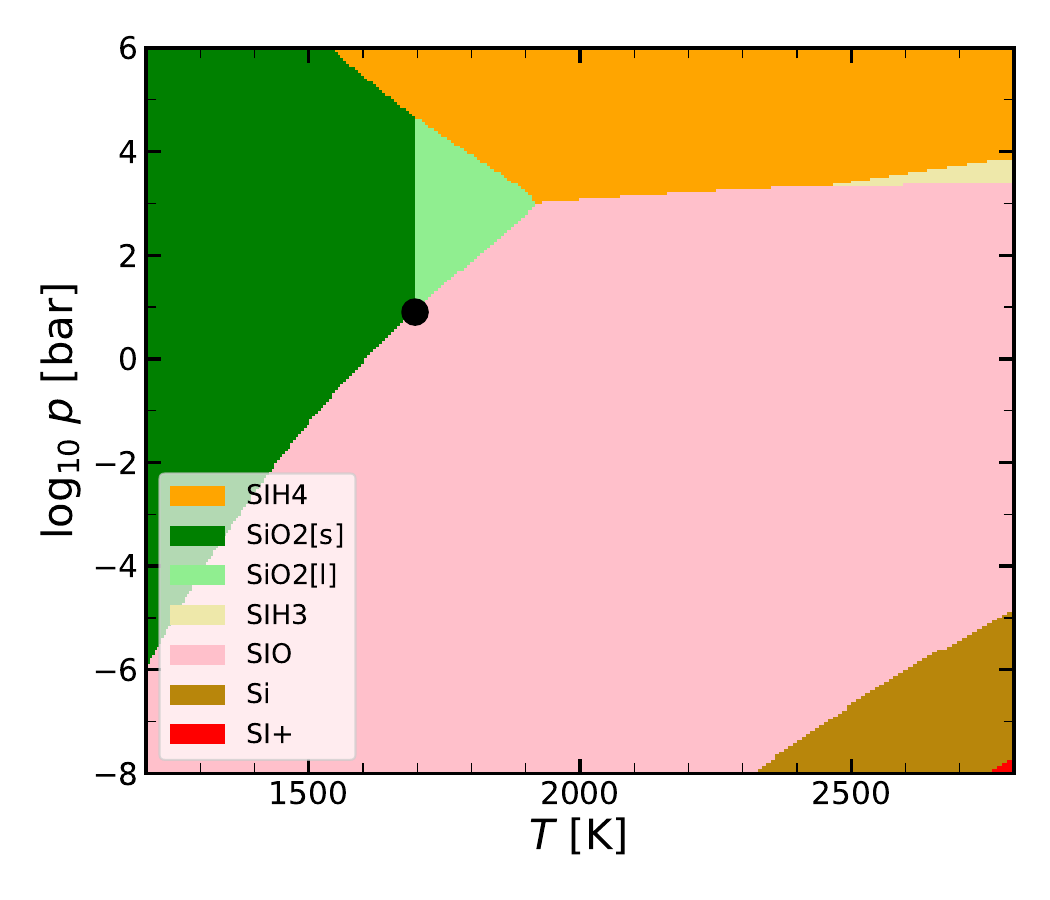} &
  \includegraphics[page=3,width=80mm,trim=0 0 0 0,clip]
                  {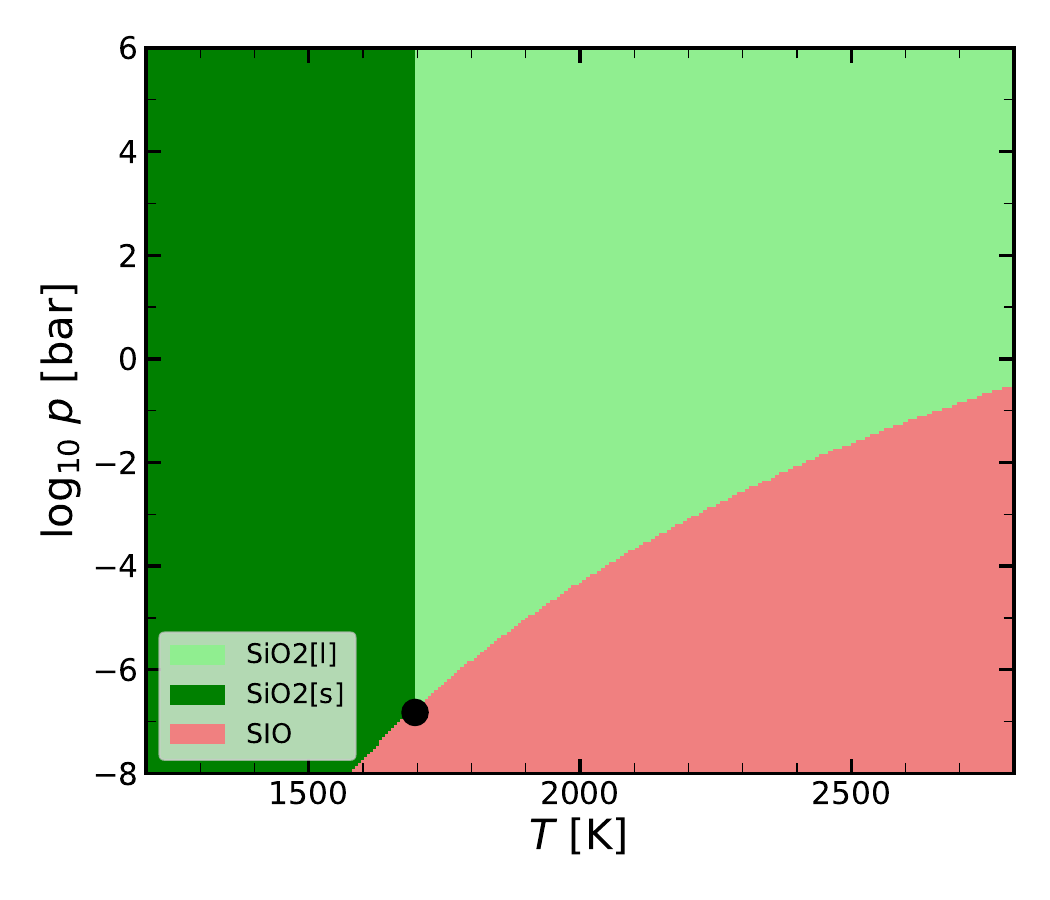}\\*[-40mm]
  \begin{minipage}{6mm}{\sf Al$_2$O$_3$}:\\*[70mm]\end{minipage} &
  \includegraphics[page=4,width=80mm,trim=0 0 0 0,clip]
                  {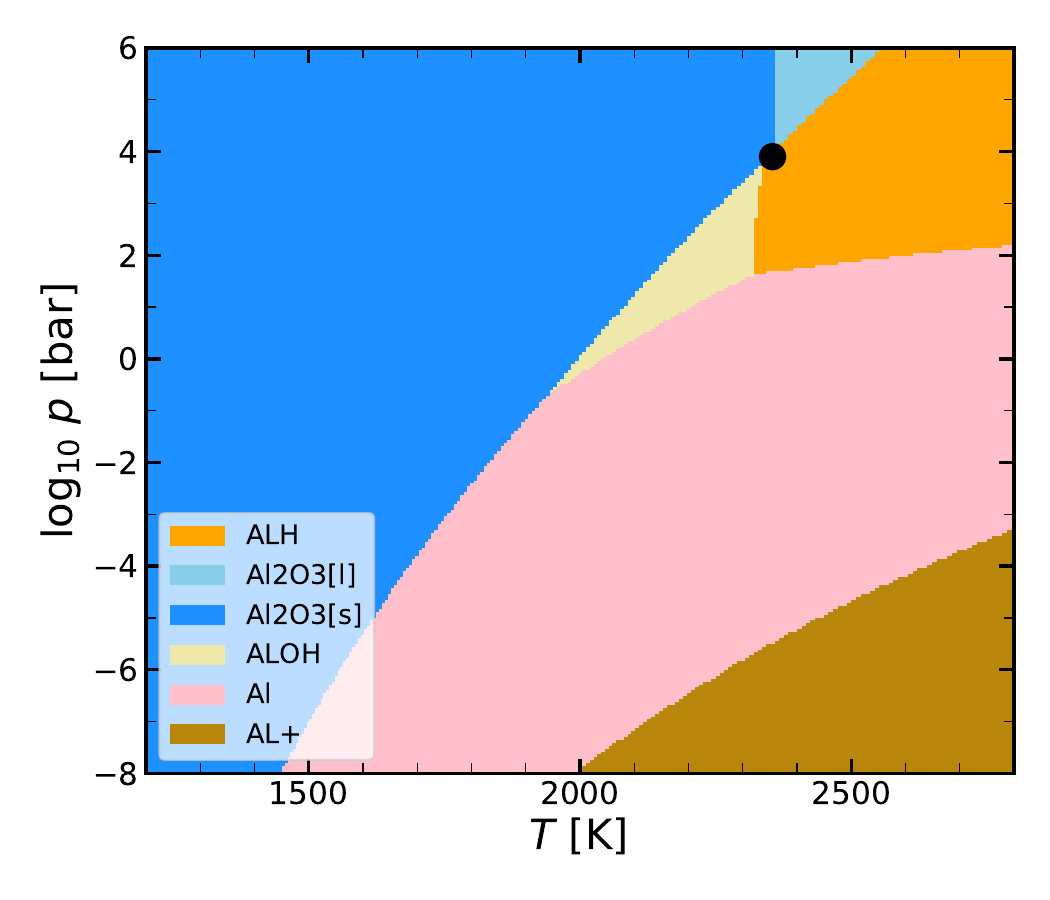} &
  \includegraphics[page=4,width=80mm,trim=0 0 0 0,clip]
                  {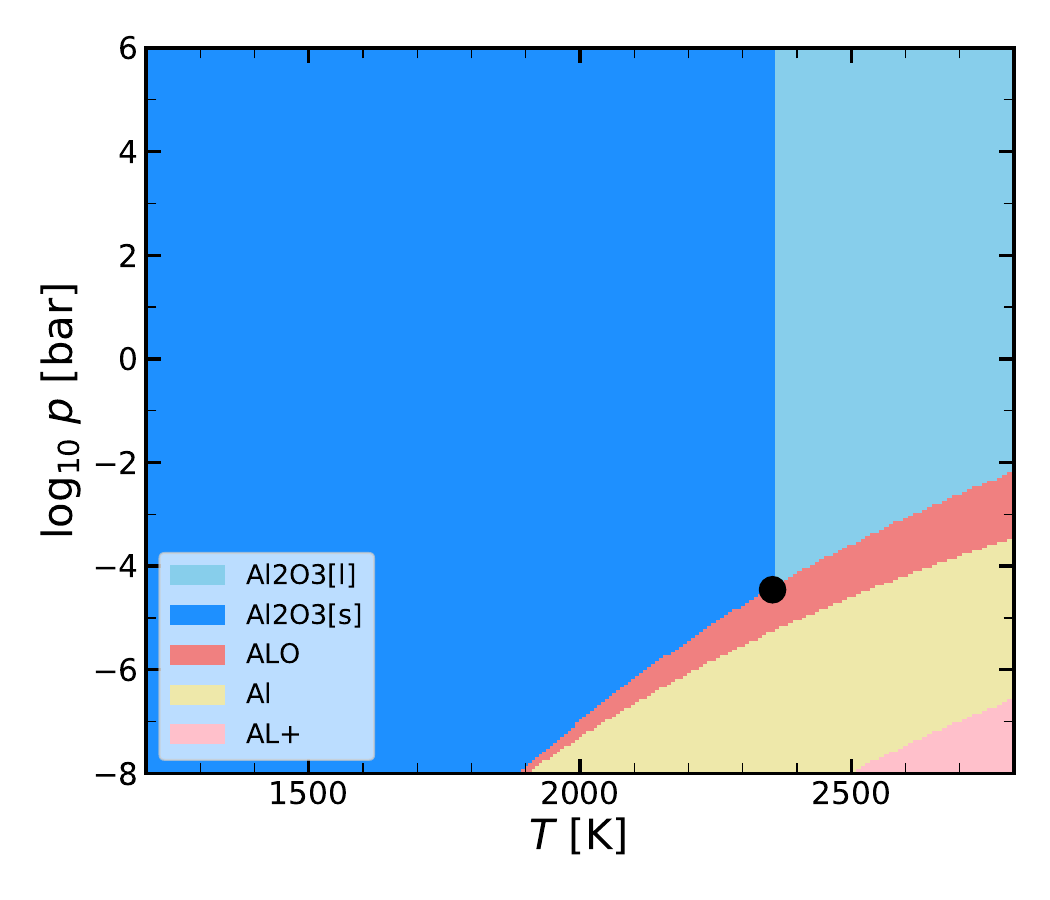}\\*[-40mm]
  \end{tabular}    
  \caption{Phasediagrams for quartz (\ce{SiO2}) and corundum (\ce{Al2O3}) in H-rich (left) and H-poor (right) environments.  The black circles mark the triplepoints of quartz and corundum in these cases. }
  \label{fig:phasedia}
\end{figure*}

In order to discuss the availability of liquid phases in a protoplanetary disc, we performed simple GGchem simulations with only 4 elements (H, O, Si, Al) for hydrogen-rich and hydrogen-poor environments.  For the hydrogen-rich model, we used solar abundances \citep{Asplund2009}: $\rm 
\epsilon_{H}\!=\!12, 
\epsilon_{O}\!=\!8.69,
\epsilon_{Al}\!=\!6.45,
\epsilon_{Si}\!=\!7.51$.
For the hydrogen-poor model we used $\rm \epsilon_{H}\!=\!1$ instead while leaving the other element abundances unchanged. The resulting phasediagrams are shown in Fig.~\ref{fig:phasedia}.  The coloured areas show the regions in the $(p,T)$-plane where a certain molecule, solid [s] or liquid [l] contains most of an element.  We read off the triplepoints for quartz in hydrogen-rich and hydrogen-poor environments as $\rm (8\,bar, 1696\,K)$ and $\rm (1.5\!\times\!10^{-7}\,bar,1696\,K)$, respectively.  For corundum, we find $\rm (7000\,bar,2327\,K)$ and $\rm (2.5\times10^{-5}\,bar,2327\,K)$, respectively. 

In the H-rich case, \ce{H2} is the dominating molecule, providing the gas pressure, and most of the oxygen is locked in \ce{H2O}.  Both effects reduce the stability and amounts of Si-oxide and Al-oxide molecules in the gas phase required for the condensation of quartz and corundum, and hence large gas pressures are required for condensation. In the H-poor case, the most abundant molecule is \ce{O2}, providing the gas pressure, and Si-oxide and Al-oxide molecules form easier and in larger quantities.  In that case, condensates can form already at much lower gas pressures -- the difference is about 8\,orders of magnitude in pressure.  This finding has been described as "chemical sputtering" by \cite{Gail1999}:  At the surface of a hot oxide grain, the collisions with \ce{H2}-molecules and \ce{H2O} tend to destabilise and decompose the refractory materials; oxides are less stable in a reducing atmosphere.  At millibar pressures, the solids decompose in the presence of \ce{H2} and sublimate before the melting point is reached.

The H-poor case resembles most situations studied in laboratory experiments or observations on Earth, where it is possible to create a melt of a refractory material by heating. However, gas pressures as high as 8\,bar and 7000\,bar (for quartz and corundum, respectively) are required for the existence of liquid phases in a solar composition gas. Such pressures are not available in protoplanetary discs, where the gas pressures barely reach a few millibars in the closest midplane during disc evolution. Therefore, our conclusion is that liquid phases are not available during disc evolution.  A true melting of oxide materials only seems possible in the interior of gravitationally bound bodies such as planetesimals, where such pressures can be reached, and where the melt is not in direct contact with \ce{H2}.

\end{document}